\documentclass[amsmath,amssymb,aps,twocolumn,pra,freq,floatfix, superscriptaddress]{revtex4-2}
\usepackage{graphicx}
\usepackage{dcolumn}
\usepackage{bm}
\usepackage{hyperref}
\usepackage{color}
\usepackage{subfigure}
\usepackage{flexisym}
\usepackage{float}
\usepackage{placeins}

\begin{document}

\title{Fiber-coupled Diamond Magnetometry with an Unshielded 30 pT/$\sqrt{\textrm{Hz}}$ Sensitivity}


\author{S. M. Graham}
\altaffiliation{s.graham.4@warwick.ac.uk}
\affiliation{Department of Physics University of Warwick Gibbet Hill Road Coventry CV4 7AL United Kingdom}
\affiliation{Diamond Science and Technology Centre for Doctoral Training University of Warwick Coventry CV4 7AL United  Kingdom}%

\author{A. T. M. A. Rahman}
\affiliation{Department of Physics University of Warwick Gibbet Hill Road Coventry CV4 7AL United Kingdom}
 

 \author{L. Munn}
 \affiliation{Department of Physics University of Warwick Gibbet Hill Road Coventry CV4 7AL United Kingdom}
 

\author{R. L. Patel}
\affiliation{Department of Physics University of Warwick Gibbet Hill Road Coventry CV4 7AL United Kingdom}
\affiliation{Diamond Science and Technology Centre for Doctoral Training University of Warwick Coventry CV4 7AL United  Kingdom}%
 

 \author{A. J. Newman}
 \affiliation{Department of Physics University of Warwick Gibbet Hill Road Coventry CV4 7AL United Kingdom}
 \affiliation{Diamond Science and Technology Centre for Doctoral Training University of Warwick Coventry CV4 7AL United Kingdom}%

\author{C. J. Stephen}
\affiliation{Department of Physics University of Warwick Gibbet Hill Road Coventry CV4 7AL United Kingdom}

\author{G. Colston}
\affiliation{Department of Physics University of Warwick Gibbet Hill Road Coventry CV4 7AL United Kingdom}

 \author{A. Nikitin}
 \affiliation{Department of Physics University of Warwick Gibbet Hill Road Coventry CV4 7AL United Kingdom}

\author{A. M. Edmonds}
\affiliation{Element Six Innovation Fermi  Avenue Harwell  Oxford Didcot  OX11 0QR  Oxfordshire  United  Kingdom}%

\author{D. J. Twitchen}
\affiliation{Element Six Innovation Fermi  Avenue Harwell  Oxford Didcot  OX11 0QR  Oxfordshire  United  Kingdom}%

\author{M. L. Markham}
\affiliation{Element Six Innovation Fermi  Avenue Harwell  Oxford Didcot  OX11 0QR  Oxfordshire  United  Kingdom}%

\author{G. W. Morley}
\altaffiliation{gavin.morley@warwick.ac.uk}
\affiliation{Department of Physics University of Warwick Gibbet Hill Road Coventry CV4 7AL United Kingdom}%
\affiliation{Diamond Science and Technology Centre for Doctoral Training University of Warwick Coventry CV4 7AL United Kingdom}%


%


\begin{abstract}
Ensembles of nitrogen vacancy centres (NVCs) in diamond can be employed for sensitive magnetometry. In this work we present a fiber-coupled NVC magnetometer with an unshielded sensitivity of (30 $\pm$ 10) pT/$\sqrt{\textrm{Hz}}$ in a (10 - 500)-Hz frequency range. This sensitivity is enabled by a relatively high green-to-red photon conversion efficiency, the use of a [100] bias field alignment, microwave and lock-in amplifier (LIA) parameter optimisation, as well as a balanced hyperfine excitation scheme. Furthermore, a silicon carbide (SiC) heat spreader is used for microwave delivery, alongside low-strain 1 $\textrm{mm}^3$ $^{12}\textrm{C}$ diamonds, one of which is placed in a second magnetically insensitive fluorescence collecting sensor head for common-mode noise cancellation. 
The magnetometer is capable of detecting signals from sources such as a vacuum pump up to 2 m away, with some orientation dependence but no complete dead zones, demonstrating its potential for use in remote sensing applications. 

\end{abstract}

\maketitle

\section{Introduction}

High sensitivity magnetometry has many applications, from fundamental physics \cite{bouchard2011detection, acosta2019color, dovzhenko2018magnetostatic, afach2021search} to biology \cite{steinert2013magnetic, glenn2015single, barry2016optical, davis2018mapping}, geoscience \cite{glenn2017micrometer, gooneratne2017downhole, ingleby2022digital} and industrial sensing \cite{zhang2021battery, zhou2021imaging, hatano2021simultaneous, chatzidrosos2019eddy}. Over the past decade there has been considerable interest in using solid-state defects for magnetometry, with the negative charge state of the NVC in diamond being the most promising candidate \cite{taylor2008high, acosta2009diamonds}. NVC magnetometers do not require heating, cryogenic cooling, vacuum tubes or magnetic shielding. Ensemble NVC magnetometers can obtain millimeter-scale spatial resolution with high dynamic range \cite{taylor2008high, rondin2014magnetometry, barry2020sensitivity, zhou2021imaging}. The properties of diamond and the NVC make them suitable for extreme environments, NVCs being capable of sensing at temperatures up to 600 K \cite{plakhotnik2014all, toyli2013fluorescence, toyli2012measurement, liu2019coherent}  and pressures of 60 GPa \cite{doherty2014electronic, ivady2014pressure, hsieh2019imaging}, as well as for biological and medical applications owing to the low toxicity of diamond \cite{barry2020sensitivity, zhu2012biocompatibility}. The use of ensembles of NVCs increases sensitivity by a factor of 1/$\sqrt{\textrm{N}}$, where N is the number of sensing spins, at the expense of spatial resolution \cite{acosta2009diamonds, taylor2008high, rondin2014magnetometry, barry2020sensitivity}. Ensembles also enable vector magnetometry, without the need for multiple sensor heads \cite{le2013optical, maertz2010vector, pham2011magnetic, schloss2018simultaneous, steinert2010high, barry2020sensitivity, dale2017medical}. 

Most high-sensitivity NVC magnetometers are entirely confined to table tops, with more portable devices having significantly inferior sensitivity. To improve the functionality
of NVC magnetometers, the optoelectronic equipment can be fiber coupled to a small mobile sensor head \cite{rabeau2005diamond, liu2013fiber, ampem2009nano, mayer2016direct, ruan2015nanodiamond, henderson2011diamond, liu2013fiber, fedotov2014fiber, fedotov2014fiber2}. 
In recent years several fiber-coupled NVC magnetometers with subnanotesla sensitivities have been demonstrated \cite{patel2020subnanotesla, sturner2021integrated, chatzidrosos2021fiberized}. The use of a tracking pulsed magnetometry scheme has allowed a sensitivity of 103 pT/$\sqrt{\textrm{Hz}}$ in a (1 - 2600)-Hz frequency range to be achieved, setting the current record for fiber-coupled NVC magnetometers \cite{zhang2022high}. This may be compared to the records of $\break$ 0.9 pT/$\sqrt{\textrm{Hz}}$ and 15 pT/$\sqrt{\textrm{Hz}}$ for non-fiber-coupled NVC magnetometers with and without ferrite flux concentrators respectively \cite{fescenko2020diamond, barry2016optical}. A magnetically shielded room has been used for the 15-pT/$\sqrt{\textrm{Hz}}$ magnetometer \cite{barry2016optical}. High-sensitivity NVC magnetometers with portable fiber-coupled sensor heads are more suitable than purely table-top setups for a wide range of applications, from magnetocardiography (MCG) \cite{dale2017medical, arai2022millimetre} to the detection of corrosion in steel \cite{chatzidrosos2019eddy, zhou2021imaging}.

In this work, we demonstrate a fiber-coupled NVC magnetometer with a sensitivity of 30 pT/$\sqrt{\textrm{Hz}}$ in a $\break$ (10 - 500)-Hz frequency range. 
The ability of the magnetometer to remotely detect real-life magnetic signals is demonstrated here using an Edwards Vacuum TIC vacuum pump. The vacuum pump consists of a backing pump, the magnetic signals from which can be seen up to 2 m away, and a turbo pump, with a 1-kHz magnetic signal that can be observed at distances up to 1.8 m.

A continuous-wave optically detected magnetic resonance (cw ODMR) magnetometry scheme, in which the NVC ensemble is addressed simultaneously and continuously with both microwaves and a laser is employed here. A bias magnetic field is applied to split out the magnetic resonances. A LIA is used to pick out the magnetometry signal from the noise and a fixed microwave frequency is applied to the steepest part of a given resonance, the zero-crossing point. A magnetic field from a sample will shift the frequency of the resonance via the Zeeman effect. This leads to changes in the NVC fluorescence, and thus shifts in the LIA voltage output. Using a calibration constant, the magnetic field responsible for the voltage shift can be determined in real time, assuming that the shift is within the bandwidth and dynamic range of the magnetometer \cite{levine2019principles}. The physics of the NVC is explained in appendix A. 

For cw ODMR the fundamental photon-shot-noise-limited sensitivity, $\eta$ is given by \cite{dreau2011avoiding, barry2016optical}

\begin{equation}
\label{eq:PhotonShotNoiseLimit}
\eta = \frac{4}{3\sqrt{3}}\frac{\Delta\nu}{\gamma C\sqrt{R}},
\end{equation}

where $\Delta$$\nu$ is the line width, $\textit{C}$ is the measurement contrast, $\textit{R}$ the photon detection rate and $\gamma$ is the gyromagnetic ratio of the NVC, equal to 28.024 GHz/T.

As seen in Eq. (\ref{eq:PhotonShotNoiseLimit}) optimization of the sensitivity requires the maximization of $\textit{R}$, and thus the fluorescence collection from the NVCs, as well as the C/$\Delta$$\nu$ ratio. The first of these tasks is made nontrivial by the high refractive index of diamond \cite{barry2020sensitivity}, as well as the use of a fiber. At the same time, difficulties obtaining high microwave and bias-field homogeneity over large ensemble sensing volumes, as well as microwave and laser-power broadening complicate improvements to the C/$\Delta$$\nu$ ratio. The reduction of technical noise, e.g., from laser-intensity or frequency variations, is also important for maximizing sensitivity, as in practice it is typically limited by such noise. Simply increasing the laser power to maximize fluorescence is undesirable, not only due to power broadening of the ODMR line width, but also because it leads to increased temperature fluctuations. These fluctuations change the zero-field splitting of the NVC and thus produce noise \cite{acosta2010temperature}. Very high laser powers also increase the photoionization of the NVCs \cite{wee2007two, hopper2018amplified}.

\FloatBarrier
 
\section{Methods}

\begin{figure}[t]
\includegraphics[width=\columnwidth]{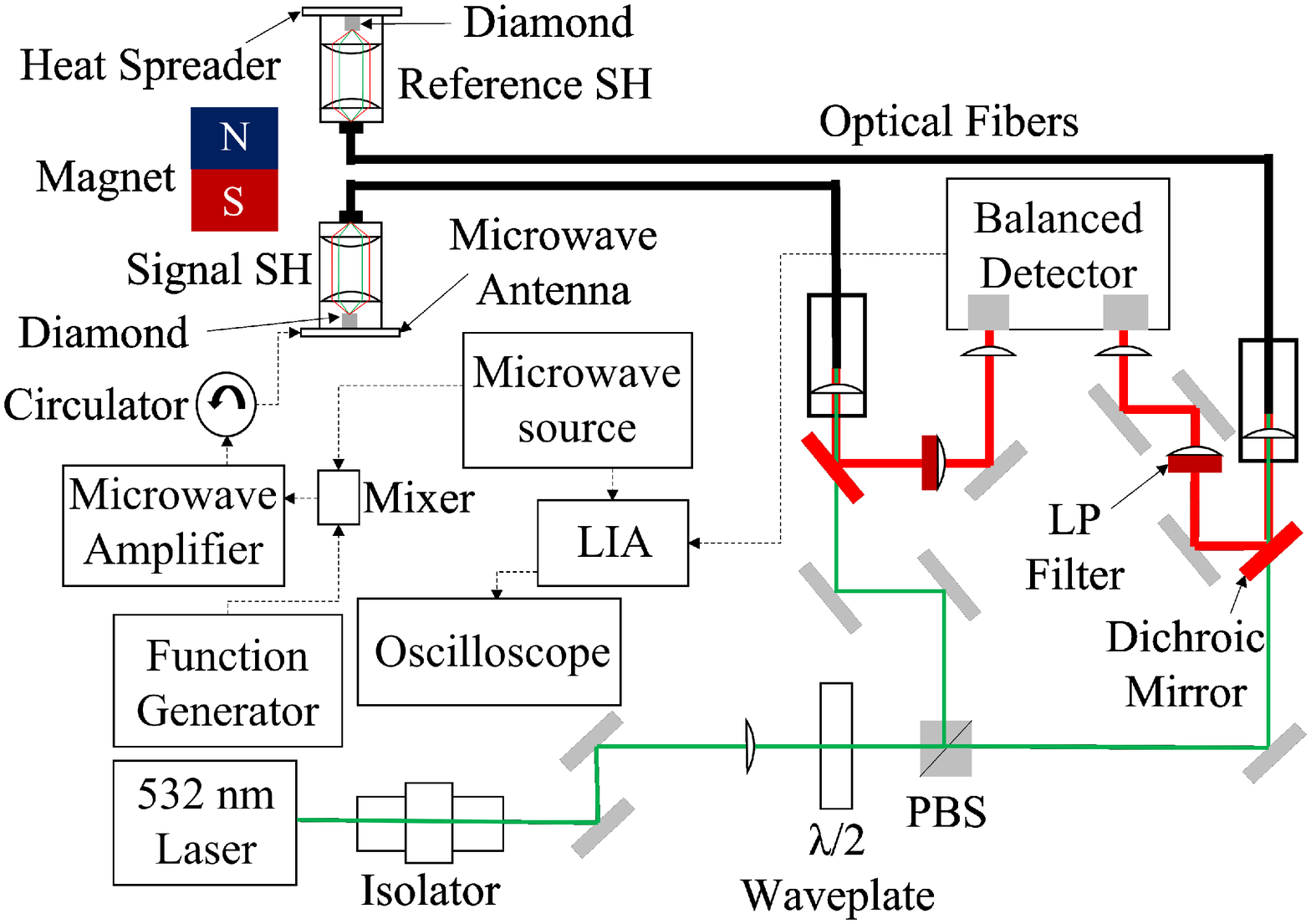} 
\caption{\small A schematic of the experimental setup: PBS, polarising beam splitter; SH, sensor head; LP, long-pass filter; LIA, lock-in amplifier.} 
\label{fig: ExperimentalSetup}
\end{figure}

The cw ODMR scheme outlined above is implemented using the two-sensor-head (signal and reference) setup shown in Fig. \ref{fig: ExperimentalSetup}. The NVC ensemble of the diamond of each sensor head is excited using a Laser Quantum 532-nm Opus laser with a maximum output of $\break$ 6 W, but to prevent saturation of the photodiodes only 1.8 W of laser power is employed, approximately 0.3 W of which is measured at the diamond within the signal sensor head. The laser power is approximately equally split between the two sensor heads. The sensor heads and optoelectronic equipment are both placed on heavy tables, which lack air legs. For each sensor head, the laser beam is focused into custom-ordered FG910UEC fibers that are $\break$ 3 m long with 0.22 NA, core diameters of 910 $\mu$m, and steel ferrule-connector-physical-contact (FC-PC) terminations. The fibers are secured to mitigate modal noise \cite{kanada1984evaluation}. Only the signal sensor head is employed for magnetometry, with the fluorescence from the magnetically insensitive reference sensor head being focused onto the reference photodiode of a Thorlabs PD450A balanced detector to allow the cancellation of common mode, most notably laser, noise. The fluorescence signals at the photodiodes from each sensor head are equal when the signal sensor head is on a magnetically sensitive resonance. The subtracted signal from the balanced detector is sent to a Zurich MFLI - 500-kHz LIA. The LIA output is supplied to a PicoScope 5442D oscilloscope. 
Within the sensor heads the fiber output is collimated using a 0.25 NA aspheric lens and then the collimated beam is focused onto the diamond using a $\break$ 0.7 NA aspheric lens. 
The same lenses and fiber are used for collection of the fluorescence. 

For microwave excitation, an Agilent E8257D microwave source and a $\break$ 43-dB Mini-Circuits ZHL-16W-43-S+ microwave amplifier are used. The microwave frequency is sine-wave modulated. 
A Mini-Circuits ZX05-U432H-S+ up-converter is used to mix the microwaves with a $\break$ 2.158-MHz sine-wave produced using an RSPro AFG21005 arbitrary function generator, enabling the use of hyperfine excitation \cite{barry2016optical}. 
Following the up-converter the microwaves are sent to a (2-4)-GHz coaxial circulator used to reduce the reflection of microwaves. 
The microwaves are supplied via a transmission line to a 1.1-mm loop antenna, with the diamond sample placed at the center, deposited upon a SiC substrate, all held within an aluminum (Al) holder with a copper (Cu) base. 
The loop and transmission line are formed from a titanium-gold (Ti-Au) contact. 
SiC combines high thermal conductivity, making it a useful heat spreader, with a dielectric nature that allows the free passage of microwaves. No microwaves are delivered to the reference sensor head, which is purely used for optical, as opposed to magnetic, noise cancellation. A 1 $\textrm{mm}^3$ low strain, 99.995\% $^{12}\textrm{C}$ CVD (100) diamond, produced by Element Six, is employed in this work. This sample is found to have a $T_2^*$ of $\break$ 750 ns. Further details on the diamond, sensor head, optics, balanced detection and hyperfine-excitation scheme may be found in Appendixes B, C, E, G, and I.

\begin{figure}[h!]
\includegraphics[width=\columnwidth]{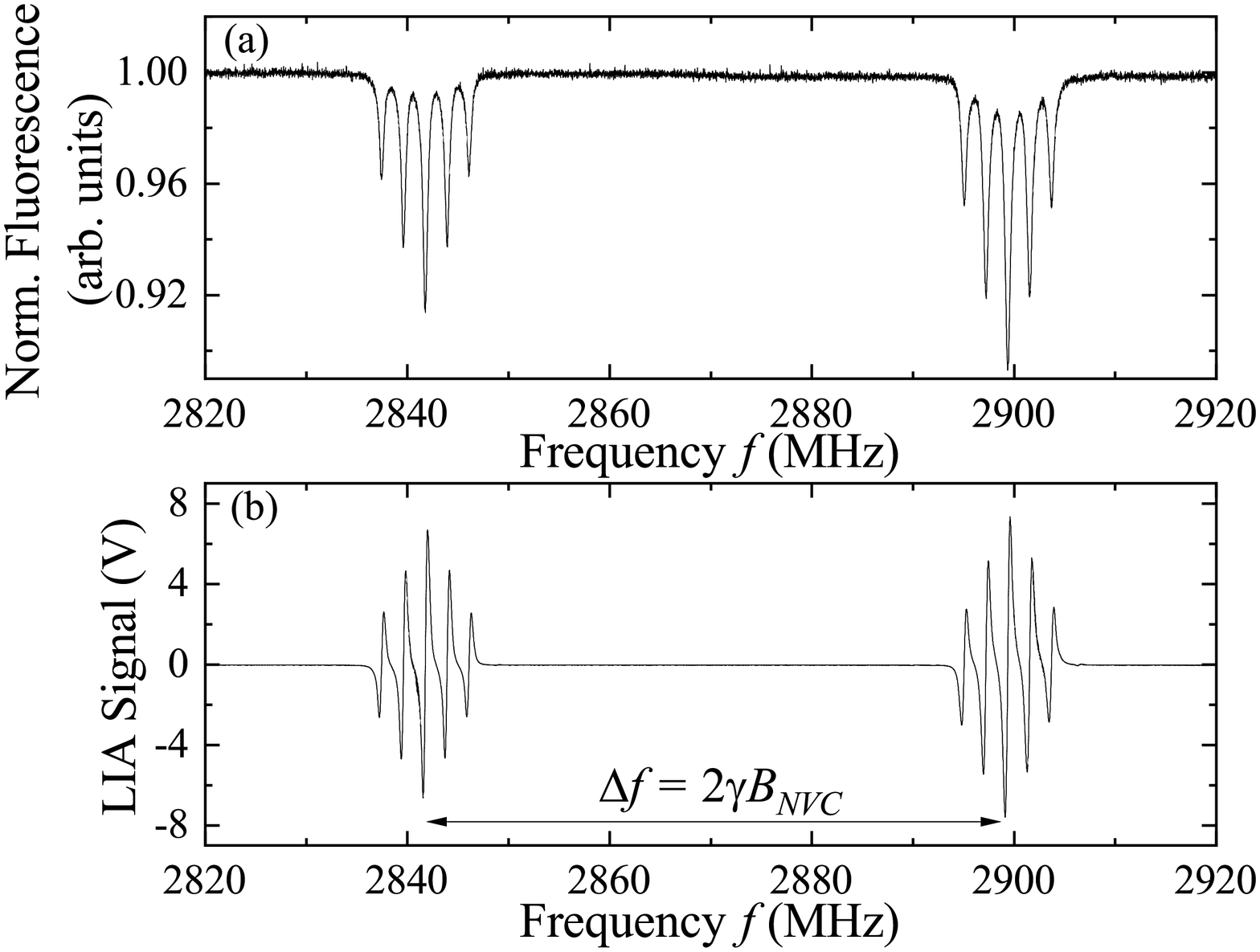} 
\includegraphics[width=\columnwidth]{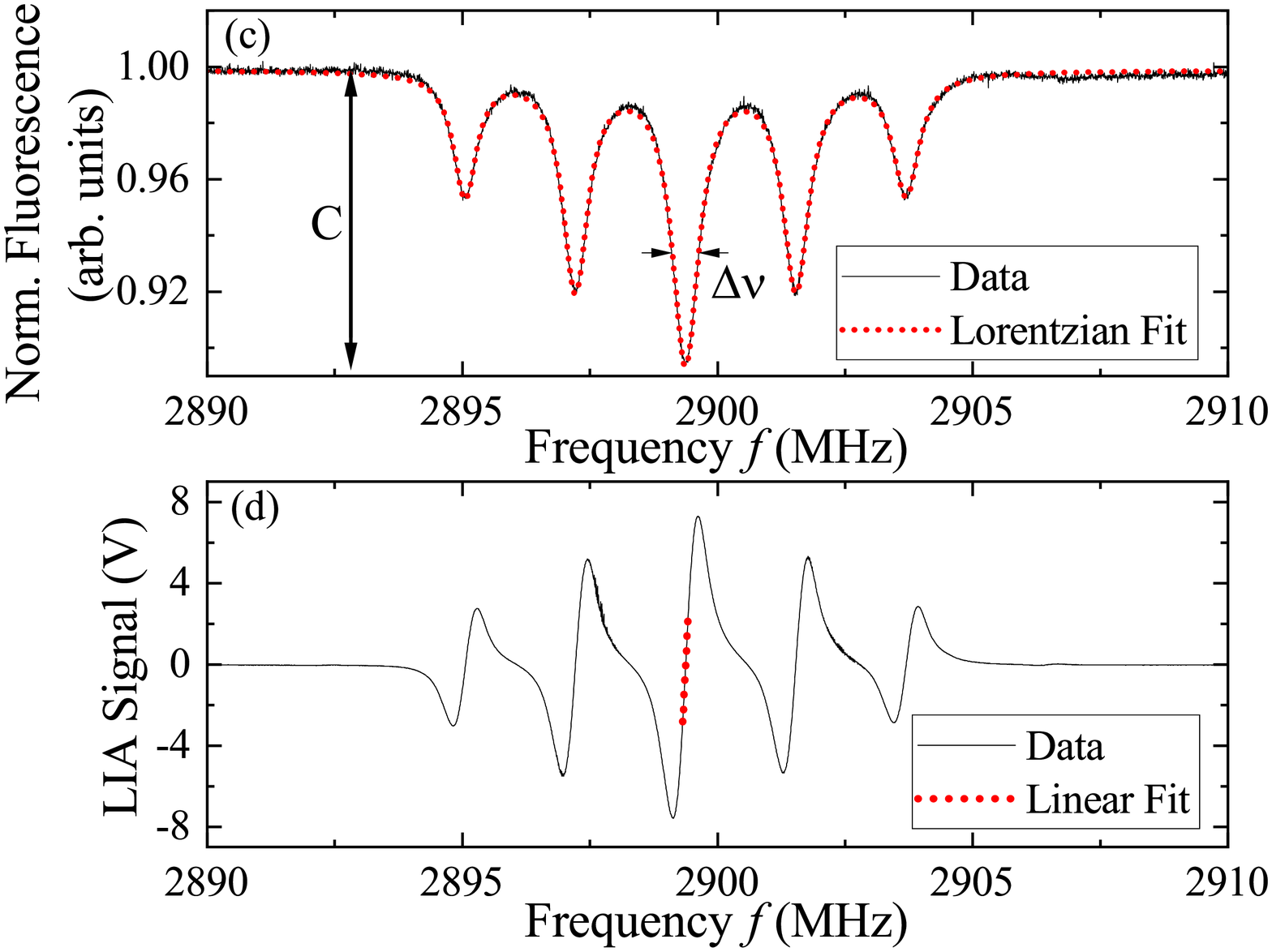} 
\caption{\small 
(a),(b) The (a) normalized ODMR and (b) demodulated ODMR spectra, respectively. $B_{NVC}$ is the projection of the external magnetic field along the NVC symmetry axis for a given alignment. (c) An enlargement of the normalized ODMR with a Lorentzian fit, with the contrast, $\textit{C}$, and line width $\Delta$$\nu$ indicated. (d) An enlargement of the demodulated ODMR spectra, with a linear fit applied at the zero-crossing point of the central resonance feature.} 
\label{fig: ODMRExample}
\end{figure}

A permanent Nd-Fe-B magnet is aligned along one of the [100] 
crystallographic orientations of the signal-sensor-head NVC ensemble. 
The use of this alignment means that the magnetometer is sensitive along a vector that projects equally onto all four possible NVC symmetry axis alignments, yielding a factor-of-4 improvement in contrast over a [111] bias-field alignment. As NVCs measure the projection of the magnetic field along their $\langle$111$\rangle$ symmetry axis and in this case the sensitive axis is not along a given NVC symmetry axis the magnetometer responsivity is reduced by an angle factor of $\break$ cos(109.5/2), which is approximately 0.58. This limits any sensitivity gain relative to a [111] alignment to a factor of 2.3 \cite{barry2016optical}. From Eq. (\ref{eq:ZeemanSplitting2}), see Fig. \ref{fig: ODMRExample}(b) and appendix A, the frequency splitting, $\Delta$$\textit{f}$, implies an applied-bias-field projection along the four possible alignments of the NVC symmetry axes equal to approximately 1 mT.

The microwave source and LIA parameters are optimized as outlined in Refs. \cite{patel2020subnanotesla, el2017optimised}. The optimum parameters with a 500-Hz LIA low-pass filter (LPF) 3-dB point, which is taken as the upper limit of the sensitive frequency range, are 3.003 kHz and 220.23 kHz for the modulation frequency and amplitude respectively. The optimum microwave power is found to occur when approximately 0.054 W is measured on a vector analyzer, placed following the microwave amplifier and circulator. 
Example ODMR spectra, including the demodulated output of the LIA, taken with these parameters are shown in Fig. \ref{fig: ODMRExample}. Parameter optimization is further discussed in appendix D. The magnetometer is, in principle, sensitive to magnetic fields at frequencies beyond the 500 Hz set by the LPF of the LIA as seen from the vacuum-pump measurements. Without this selected hardware limiter the frequency range would ultimately be set by the modulation frequency and NVC response. The 500-Hz 3-dB point was selected as this ensures that the frequency range is sufficient for observing the majority of the vacuum-pump signals, as well as for applications such as MCG. It also minimizes the magnitude of the 3.003-kHz modulation peak. For further discussion of the frequency range, see appendix J. 


\section{Results}

\begin{figure}[h!]
\includegraphics[width=\columnwidth]{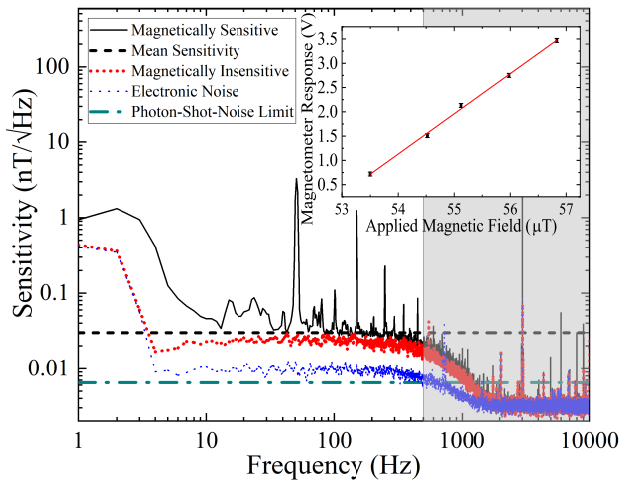} 
\caption{\small Sensitivity spectra, with a mean sensitivity of $\break$ 30 pT/$\sqrt{\textrm{Hz}}$ from 10 to 500 Hz. The noise floor is also shown when magnetically insensitive (off resonance), and with no applied laser or microwaves (electronic noise). The noise floor drops at dc due to the use of ac coupling. The shaded region lies beyond the LIA LPF 3-dB point at 500 Hz. The inset shows the magnetometer responsivity as a function of the applied magnetic test field along one of the [100] crystallographic orientations.} 
\label{fig: FFT-Sensitivity}
\end{figure}

The sensitivity is determined by applying known test fields along one of the [100] crystallographic orientations of the signal-sensor-head NVC ensemble to measure the magnetometer response. 
The test fields are produced using a Helmholtz coil calibrated using a Magnetics GM07 Hall probe. A linear fit is applied to the data seen in the inset of Fig. \ref{fig: FFT-Sensitivity} to obtain a calibration constant of $\break$ 8.3 $\times$ $10^{-4}$ V/nT. This compares to a calibration constant of 1.4 $\times$ $10^{-3}$ V/nT when using the ODMR zero-crossing slope, [see Fig. \ref{fig: ODMRExample}(d)], with the ratio of 0.60 between the two values being close to the anticipated value of 0.58. 

To obtain the noise floor, the microwave frequency is set to the zero-crossing point of the demodulated LIA ODMR spectrum. The central resonance with the largest zero-crossing slope (ZCS) is chosen. The magnetometer is then magnetically sensitive and 32 1-s time-traces are taken with a sampling rate of 20 kHz. Such time traces are also taken with the microwave frequency set off resonance: this renders the magnetometer magnetically insensitive, allowing the level of magnetic noise to be characterized by considering the difference in the noise floor between the magnetically sensitive and insensitive cases. Time traces are also taken with the laser and microwave source turned off such that the measured signal may be regarded as representing the electronic noise of the system. Using the calibration constant determined above the time traces are converted into power-spectral-density (PSD) plots before being averaged. The method of determining the sensitivity is expanded upon in appendix F. For the magnetically sensitive case the mean sensitivity is determined to be (30 $\pm$ 10) pT/$\sqrt{\textrm{Hz}}$ in a frequency range of $\break$ 10 - 500 Hz as seen in Fig. \ref{fig: FFT-Sensitivity}. When calculating this, the 50-Hz peak and its harmonics are not included, leaving the predominantly flat noise floor. 
Significant noise is present at low frequencies, below approximately 10 Hz, and this 1/$\textit{f}$ noise is attributed to laser-power variations and vibrations, as well as magnetic noise in the environment. The characteristic peaks at 50 Hz and its harmonics can be associated with ambient magnetic noise-e.g., from the mains-as they are not observed when magnetically insensitive. The average noise levels for the magnetically insensitive and electronic noise cases are $\break$ (21 $\pm$ 4) pT/$\sqrt{\textrm{Hz}}$ and (9 $\pm$ 1) pT/$\sqrt{\textrm{Hz}}$ respectively.

Equation (\ref{eq:PhotonShotNoiseLimit}) is used to calculate the photon-shot-noise limit. The red-photon detection rate is determined to be $\break$ 2.9 $\times$ $10^{15}$ Hz from the fluorescence power measured to be incident upon the balanced detector photodiodes using a Thorlabs PM100D power meter with a S121C head. The ODMR spectrum prior to LIA amplification is used to determine the line width and contrast, which are found to be 0.73 MHz and 10\%, respectively. Using these values and accounting for the 0.58 angle factor, the photon-shot-noise limit is $\break$ (6.5 $\pm$ 0.9) pT/$\sqrt{\textrm{Hz}}$. 
It would appear that NVCs distributed throughout the whole volume of the 1-$\textrm{mm}^{3}$ diamond are being addressed and contribute to the magnetometry signal. As a conservative estimate the sensing volume may thus be taken as 1 $\textrm{mm}^3$, although this may be an overestimate. This is discussed further in Appendix H.

\begin{figure}[t]
\includegraphics[width=\columnwidth]{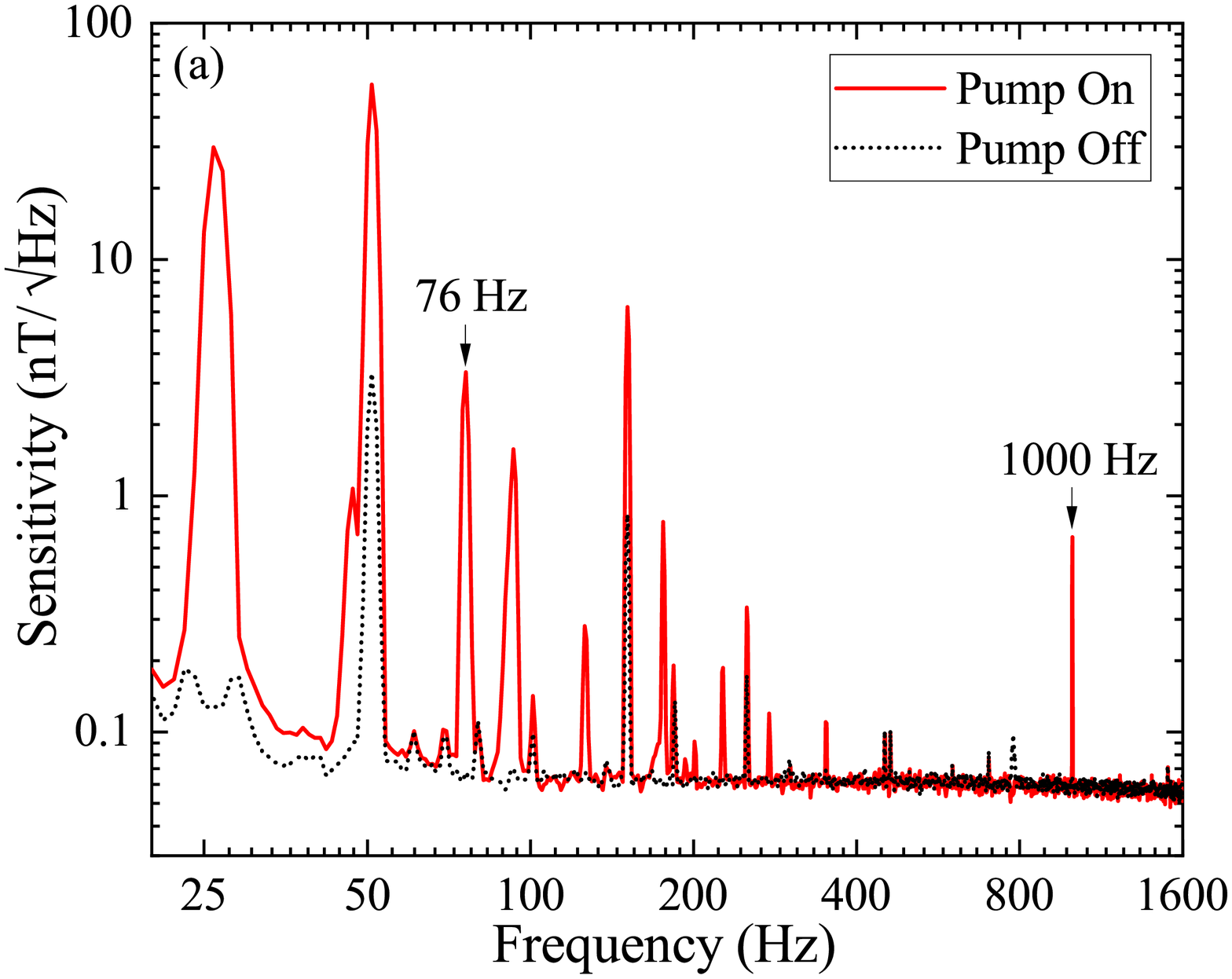} 
\includegraphics[width=\columnwidth]{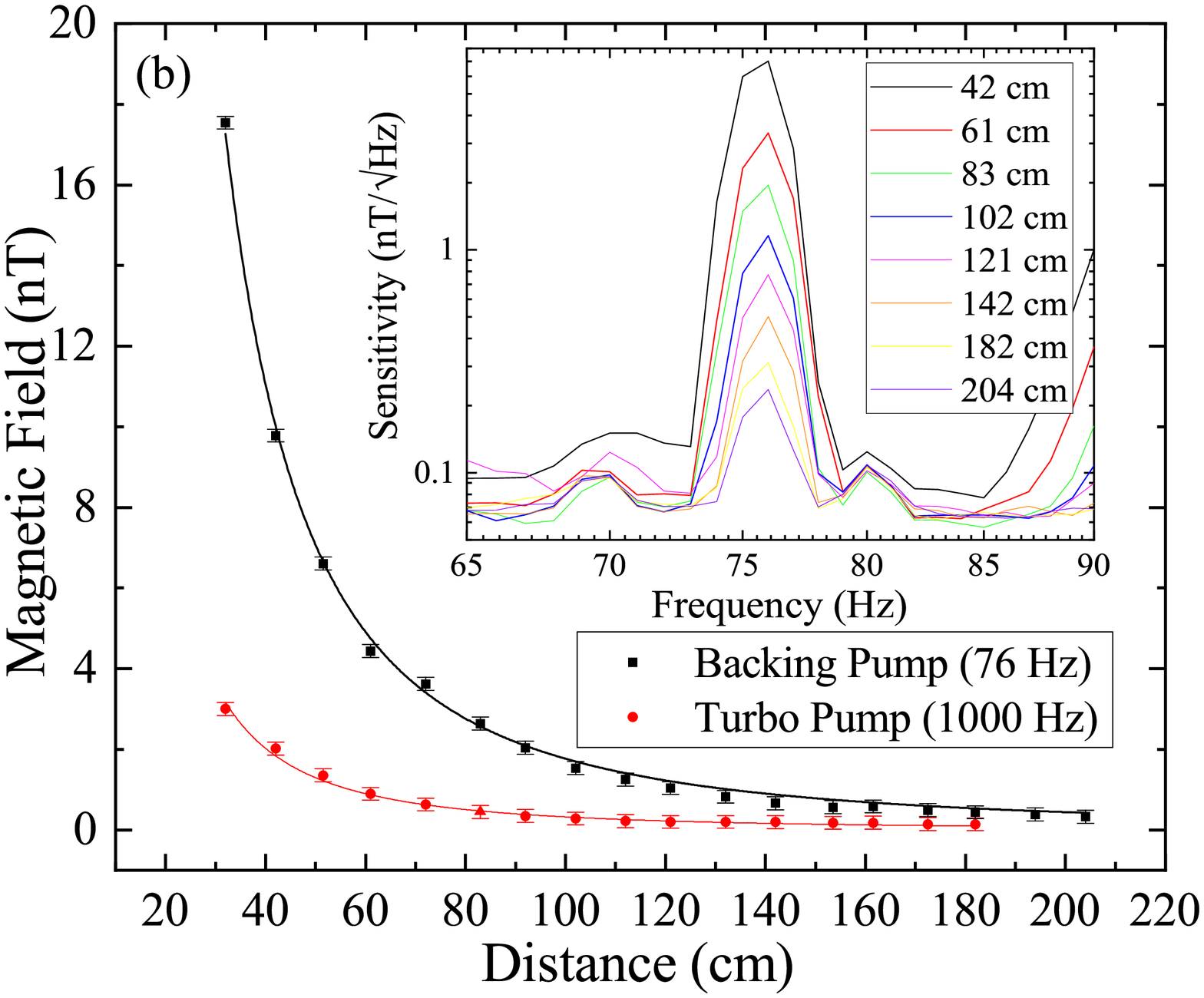}
\caption{\small (a) Sensitivity spectra taken with the vacuum pump, at a distance of 60 cm, turned on and off. 
For these measurements, the modulation frequency is set to 10.003 kHz due to the higher frequency range required. 
(b) The mean amplitudes of the 76-Hz backing-pump signal and 1-kHz turbo-pump signal as a function of the distance, both fitted with a 1/$d^2$ function. The inset shows the mean of 32 1-s fast-Fourier transforms (FFTs) of the $\break$ 76-Hz backing-pump peak.} 
\label{fig: BackingPumpOnvsOff}
\end{figure}

To demonstrate the ability of the magnetometer to detect signals from remote sources, we use a vacuum pump that consists of a backing pump and a turbo pump. An electric scooter is also investigated, as can be seen in appendix M. Figure \ref{fig: BackingPumpOnvsOff}(a) shows the magnetic signals detected from the vacuum pump compared to a reference state with the vacuum pump turned off. For these measurements, the LIA LPF 3-dB point is set to 2.8 kHz, to ensure that none of the signals are cut off. 
Averaging is used, although it is not required to observe the majority of the peaks. Many peaks can be associated with the backing pump, but we focus on a peak at 76 Hz. A peak can also be seen at 1 kHz, from the turbo pump. The amplitudes of both the backing-pump 76-Hz and 1-kHz turbo-pump signals are measured as a function of distance from 30 cm to 2 m as seen in Fig. \ref{fig: BackingPumpOnvsOff}(b), with the amplitudes following a 1/$d^2$ drop off, where d is the distance from the diamond \cite{leighton1965feynman}. 
For these measurements the vacuum pump is placed approximately level with the sensor head. The backing pump can be detected at 2 m, with the 76 Hz peak still having a signal-to-noise ratio of 5. No measurements are taken beyond 2 m due to space constraints. The turbo-pump signal can be seen at distances up to 1.8 m. 
The sensitive axis of the magnetometer lies along a [100] crystallographic orientation. Accordingly the ability to detect signals depends not only on the distance but also the orientation of the object relative to this axis. However, as there will always be a non-zero-field projection along at least one of the four NVC alignments, there is not a complete dead zone \cite{kimball_alexandrov_budker_2013} orthogonal to the sensitive axis as in Ref., \cite{patel2020subnanotesla} where a [111] bias-field alignment has been used. However, the projection and hence the response will differ for each NVC alignment if the applied fields are not aligned along [100]. This leads to a complicated overall magnetometer response and the overall response and thus sensitivity is reduced. 
This is discussed further in appendix L.

\FloatBarrier

\section{Discussion and Conclusions}

Given our contrast, line width, and fluorescence intensity, we obtain a photon-shot-noise-limited sensitivity of 6.5 pT/$\sqrt{\textrm{Hz}}$, improving upon Ref. \cite{patel2020subnanotesla} and approaching Ref. \cite{barry2016optical}. 
A relatively high ensemble contrast of 4\% is obtained by aligning the bias field along a [100] crystallographic orientation. This ensures that the entire NVC ensemble population contributes to the magnetometry signal, leading to a factor-of-4 enhancement in contrast.  
Hyperfine excitation is then found to yield a factor-of-2.5 increase in contrast, producing a high final contrast of 10\% \cite{patel2020subnanotesla, barry2016optical,sturner2019compact, sturner2021integrated}. The power is balanced between the function generator and microwave source to ensure approximately equal microwave intensity within the three hyperfine-excitation tones. The use of a SiC, as opposed to Al, antenna substrate likely results in improved microwave power delivery efficiency owing to its superior microwave transmission properties \cite{li2020experimental, patel2020subnanotesla}. The SiC may have enhance our contrast and microwave field homogeneity relative to Al \cite{patel2020subnanotesla}. The high thermal conductivity of SiC also helps to reduce noise caused by temperature fluctuations within the diamond that change the zero-field splitting of the NVC \cite{acosta2010temperature}.

The green-to-red photon conversion efficiency is approximately 0.36\%. The use of a $\break$ 910-$\mu$m-diameter fiber and aspheric lenses for the collimation and focusing of both the laser and fluorescence light enables this conversion efficiency, which is relatively high for a fiber-coupled device \cite{sturner2019compact, sturner2021integrated}, although higher efficiencies have been obtained \cite{chatzidrosos2021fiberized}. 
A comparatively narrow ODMR line width, given the NVC concentration, of 0.73 MHz is obtained \cite{patel2020subnanotesla}. 
The line width appears to be modulation-amplitude limited as without frequency modulation the line width is found to be 0.63 MHz. Compared to a $T_2^*$-limited line width of 0.42 MHz, however, it is clear that the line width also suffers from inhomogeneities in the bias magnetic field as well as microwave and laser-power broadening at the microwave and laser powers employed. Further parameter optimization is likely possible.

Furthermore, the use of a second sensor head for common-mode noise cancellation provides a factor-of-11 improvement in sensitivity compared to when using the output from a single photodiode (without balancing with picked-off green laser light). In principle, the use of a second fluorescence-emitting diamond, as opposed to just picking off a portion of the laser beam \cite{patel2020subnanotesla, sturner2021integrated, chatzidrosos2021fiberized, fescenko2020diamond}, allows for not only the laser-intensity but also laser-frequency noise to be cancelled using the balanced detector. The balanced-detection scheme is discussed further in appendix G. Nonetheless, 
we remain a factor of 4.6 above our photon-shot-noise-limited sensitivity with magnetic and laser noise being our principal limiting sources of noise, which is unsurprising in the former case given our lack of magnetic shielding. The laboratory environment is magnetically noisy with numerous colored peaks appearing, many of which may be associated with the mains.

The sensitivity of the magnetometer and the portable sensor-head design make it useful for a wide range of applications. As an example, here the machinery measurements demonstrate the capacity of NVC magnetometers for use in remote-sensing and potentially even condition-monitoring applications. In principle, the system would be capable of detecting machinery even through thick walls. Machinery such as engines, air conditioning units, and electric motors are widely used in domestic, commercial, and industrial applications. Such machinery typically produces clear and characteristic magnetic signals over a wide range of frequencies \cite{romero1996feasibility, draper2002operational}. High-sensitivity magnetometers can be used to noninvasively detect and monitor these signals and NVCs are also capable of remaining operational in harsh environments \cite{marmugi2017remote, wang2006condition, verma2020automatic, barry2020sensitivity}. If monitored in working condition for a long period of time, shifts in the frequencies of these characteristic peaks could then be used by operators to identify damage. High magnetic sensitivities enable detection even with large separations between the sensor and the target. Vector magnetometry could be used to obtain additional information and to remove the orientation dependence \cite{clevenson2018robust, schloss2018simultaneous, chatzidrosos2021fiberized}. Remote condition monitoring is also discussed, alongside measurements taken using an electric scooter, in appendix M. With further improvements in sensitivity, the magnetometer could potentially be used for MCG, for which the portable sensor head is convenient as it can be brought into close contact with the chest. The relatively small sensing volume of the diamond would allow, in principle, for high spatial resolutions and the diamond could be made smaller. For MCG, the reduction of 1/$\textit{f}$ noise would be crucial, given that the magnetic signals of the heart are to be found in the range $\break$ 1 - 40 Hz \cite{dale2017medical, fenici2005clinical, chatzidrosos2017miniature, jensen2018magnetocardiography, arai2022millimetre}. Many sensor heads could be employed to allow for rapid mapping of the entire heart \cite{yang2021new}. Magnetoencephalography (MEG) would be a possible application if femtotesla sensitivities could be obtained \cite{cohen1972magnetoencephalography, boto2018moving}.  Other possible applications, achievable with present sensitivities, include battery monitoring \cite{hatano2022high, hu2020sensitive}.

The setup could be further improved via gradiometry, which could be implemented by delivering microwaves to the reference sensor head. This should allow magnetic noise to be cancelled, assuming that it is common to both sensor heads. This would be especially effective at reducing the impact of colored mains noise, such as the 50-Hz peak, and potentially also the 1/$\textit{f}$ noise. 
Gradiometry has been employed in several other NVC magnetometers \cite{blakley2015room, blakley2016fiber, masuyama2021gradiometer, zhang2021diamond} and the possibility of implementing gradiometry for our setup is expanded upon in appendix G.

The relatively low photon-collection efficiency, compared to non-fiber-coupled devices, remains an issue \cite{wolf2015subpicotesla, ma2018magnetometry}. 
Directly coupling the fiber to the diamond \cite{zhang2022high} or placing the diamond within the fiber may allow for improvements in photon-collection efficiency. This would also allow for reductions in the size of the sensor head, allowing it to be brought closer to objects of interest. Hollow-core fibers could be useful for this, although such fibers that are commercially available typically have relatively high losses \cite{maayani2019distributed, shibata1981prediction, lines1984search, sakr2020hollow}. Reflective surface coatings, as well as antireflection coatings, have also been shown to improve fluorescence collection and excitation efficiency \cite{yeung2012anti, zhang2022high}. 

Ferrite flux concentrators have been employed in several NVC magnetometers to concentrate magnetic flux from samples into a diamond, yielding up to a factor-of-254 improvement in sensitivity \cite{fescenko2020diamond, xie2021hybrid}. However, they limit spatial resolution and it would be relatively difficult to implement ferrite flux concentrators with our current sensor-head design. Dual-resonance magnetometry could also be employed to ensure that the NVC magnetometer is less susceptible to noise, caused by temperature fluctuations, which is problematic for the target applications \cite{acosta2010temperature, clevenson2015broadband, schloss2018simultaneous, clevenson2018robust, fescenko2020diamond}.

Pulsed schemes typically yield around an order-of-magnitude improvement in sensitivity over cw schemes for a given excitation volume \cite{barry2016optical, dreau2011avoiding, barry2020sensitivity}, while also enabling the implementation of spin-bath driving and double quantum magnetometry techniques \cite{bauch2018ultralong, knowles2014observing}. However, high microwave and optical excitation homogeneity is critical for fully exploiting pulsed schemes and it would likely be necessary to use a microwave resonator to efficiently drive all the NVCs with comparable dynamics, which would limit the bandwidth of our NVC magnetometer \cite{chen2018large, eisenach2018broadband, ahmadi2017pump, bayat2014efficient}. A more uniform intensity could be obtained across the excitation volume using a top-hat beam shaper to convert the Gaussian laser output to be super-Gaussian \cite{lu2002study, cherezova1998super}. Fully exploiting the benefits of pulsed magnetometry would also require the saturation of the NVC ensemble, and thus significantly higher laser power for the current diamond sample. Alternatively, a smaller diamond, with a correspondingly smaller NVC ensemble, would be more readily saturated at lower laser powers. The use of a smaller diamond sample would also help with obtaining high microwave and bias-field homogeneity. However, a smaller NVC ensemble would lead to a lower photon detection rate, $\textit{R}$. For further discussion of the saturation behavior of the NVCs, see Appendix H. 

In conclusion, we present a fiber-coupled NVC magnetometer with an unshielded sensitivity of $\break$ (30 $\pm$ 10) pT/$\sqrt{\textrm{Hz}}$ in a (10 - 500)-Hz frequency range with a sensing volume of approximately 1 $\textrm{mm}^3$ or less. It is shown to be capable of remotely detecting real-life magnetic fields from machinery. The portable sensor head combined with the high sensitivity make it useful for a wide range of applications. All the data used in the production of this work is available online \cite{SMGraham2023}. 

\section{Acknowledgements}

We thank Jeanette Chattaway, Matty Mills and Lance Fawcett of the Warwick Physics Mechanical workshop, and Robert Day and David Greenshield of the Warwick Physics electronics workshop. The Ph.D. studentships of S. M. G. and A. J. N. are funded by the Defence Science and Technology Laboratory (DSTL) and the National Nuclear Laboratory (NNL) respectively. R. L. P's Ph.D. studentship is funded by the UK Engineering and Physical Sciences Research Council (EPSRC) Centre for Doctoral Training in Diamond Science and Technology (Grant No. EP/L015315/1). L. M. is supported by the Warwick University Undergraduate Research Support Scheme (URSS). This work is supported by the UK Hub NQIT (Networked Quantum Information Technologies) and the UK Hub in Quantum Computing and Simulation, part of the UK National Quantum Technologies Programme, with funding from UK Research and Innovation (UKRI) EPSRC Grants No. EP/M013243/1 and No. EP/T001062/1, respectively. This work is also supported by Innovate UK Grant No. 10003146, EPSRC Grant No. EP/V056778/1 and EPSRC Impact Acceleration Account (IAA) awards (Grants No. EP/R511808/1 and No. EP/X525844/1). G.W.M. is supported by the Royal Society.

\FloatBarrier

\section*{Appendix A: Physics of the NVC}

\begin{figure}[h!]
\includegraphics[width=\columnwidth]{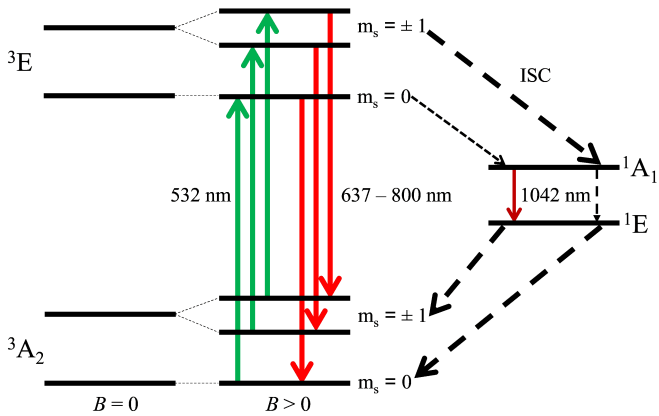} 
\caption{\small A schematic of the electronic structure of the NVC in diamond, with electric dipole transitions and intersystem crossing (ISC) indicated.} 
\label{fig: NVCentre}
\end{figure}

The NVC is a point defect in the diamond lattice consisting of a substitutional nitrogen atom adjacent to a vacancy. The defect has trigonal symmetry, with a $\langle$111$\rangle$ major symmetry axis taken as the line between the nitrogen and vacancy, which has four possible alignments [111], [$\overline{1}$$\overline{1}$1], [1$\overline{1}$$\overline{1}$], and [$\overline{1}$$1\overline{1}$] for a given diamond crystal \cite{doherty2013nitrogen, doherty2014measuring}. 
In an ensemble these alignments will typically be equally populated. The negative charge state of the NVC is a spin-1 defect with a spin-triplet ground state that can be optically initialized and read out using ODMR spectroscopy \cite{rondin2014magnetometry, barry2020sensitivity}. The energy-level structure of the NVC is shown in Fig. \ref{fig: NVCentre}. The NVC can be optically initialized into the $m_s$ = 0 sublevel of the $^3A_2$ ground state using a 532-nm laser, as if the defect is initially in the $\break$ $m_s$ = $\pm$ 1 sublevels prior to excitation, it has a greater probability of returning to the ground state via the singlet states, which involves non-spin-conserving transitions. This pathway is also nonradiative, allowing the spin state to be read out from the fluorescence intensity. The spin state can be manipulated and driven from the $m_s$ = 0 to $m_s$ = $\pm$ 1 sublevels using microwaves
\cite{rondin2014magnetometry, barry2020sensitivity}.

The zero-field splitting of the $^3A_2$ ground state is $\break$ approximately 2.87 GHz at room temperature. Upon the application of an external magnetic field the degeneracy of the $\break$ $m_s$ = $\pm$ 1 sublevels is lifted via the Zeeman effect, leading to resonances at two different frequencies (for a given NVC alignment) split by \cite{rondin2014magnetometry, barry2020sensitivity}

\begin{equation}
\label{eq:ZeemanSplitting2}
\Delta f = 2\gamma B_{NVC},
\end{equation}

where $B_{NVC}$ is the projection of the external magnetic field along the NVC symmetry axis for a given alignment. The $\break$ S = 1 electron spin has a hyperfine interaction with the $\break$ I = 1 nuclear spin of the $^{14}\textrm{N}$ atom of the NVC, splitting each resonance into three, separated by approximately 2.158 MHz \cite{doherty2013nitrogen, barry2016optical}. 

\FloatBarrier

\section*{Appendix B: Diamond Properties} 

An isotopically purified (99.995\% $^{12}\textrm{C}$) CVD diamond is used in this work, ensuring a minimal concentration of $\break$ I = 1 $^{13}\textrm{C}$ nuclear spins \cite{balasubramanian2009ultralong, edmonds2021characterisation} that would otherwise contribute to decoherence. The diamond is electron irradiated and then annealed to increase its NVC concentration, which is found to be (2.8 $\pm$ 0.2) parts per million via EPR measurements taken using a Bruker EMX spectrometer with a 90-dB microwave bridge and a Bruker superhigh-Q cavity. Pulsed EPR measurements are taken with a Bruker E580 and MD5 resonator to determine $T_2^*$, $T_2$, and $T_1$ to be 750 ns, 1.3 $\mu$s, and 5100 $\mu$s respectively. Given this $T_2^*$ and using the formula 1/$\pi$$T_2^*$ \cite{barry2020sensitivity}, a minimum ODMR resonance line width of 0.42 MHz would be anticipated, although in practice the line width is broadened by inhomogeneities in the bias magnetic field as well as microwave and laser-power broadening. 
Following this characterization the diamond sample (which initially has dimensions of 2.97 mm $\times$ 2.93 mm $\times$ 0.93 mm) is laser cut into nine equal cube pieces, each mechanically polished so that all six faces have an optical grade finish. Two of these 1 $\times$ 1 $\times$ 1 $\textrm{mm}^3$ diamond samples are employed for the signal and reference sensor heads respectively. For effective noise cancellation it is important that the diamonds used in each of the two sensor heads are as similar to each other as possible. The EPR measurements may be found in the appendix of Ref. \cite{patel2020subnanotesla}. The diamond growth, electron irradiation, annealing, and polishing are done by Element Six. 

\section*{Appendix C: Sensor Head} 

\begin{figure}[h!]
\includegraphics[width=\columnwidth]{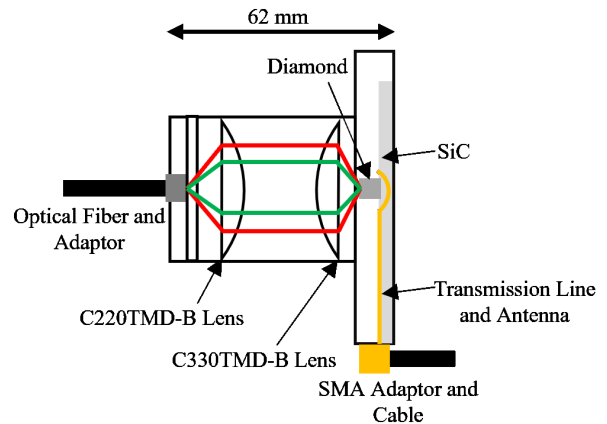} 
\caption{\small A schematic of the sensor head.} 
\label{fig: SensorHeadOptics}
\end{figure}

Figure \ref{fig: SensorHeadOptics} shows the sensor-head design in profile, including the optics, SiC substrate, and microwave antenna. 
The open 20-mm-long Thorlabs SM1L20C lens tube contains a Thorlabs $\break$ 0.25 NA C220TMD-B aspheric lens used for collimating the laser-beam output from the Thorlabs $\break$ 0.22-NA 910-$\mu$m-diameter FG910UEC fiber. The collimated laser beam is then focused down onto the 1-$\textrm{mm}^3$ diamond sample using a Thorlabs 0.7-NA C330TMD-B aspheric lens, contained within the same lens tube. This same lens combination is then used to collimate and focus the fluorescence emitted by the NVC ensemble into the optical fiber. The total length of the sensor head is 62 mm, although this could be made smaller in principle without limiting functionality. 

\begin{figure}[h!]
\includegraphics[width=\columnwidth]{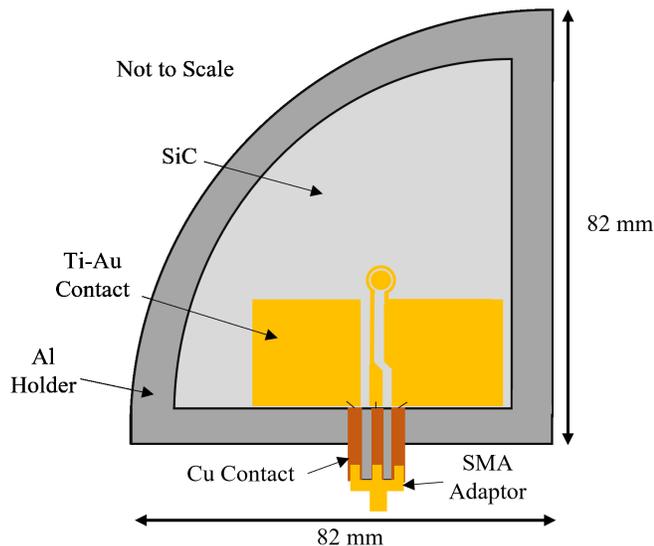} 
\caption{\small A schematic of the sensor head, showing the microwave transmission line and antenna.} 
\label{fig: SensorHeadSiCAntenna}
\end{figure}

The diamond is placed at the center of a $\break$ simple 1.1-mm loop antenna. The microwaves are supplied to this antenna by a transmission line. Initially following the SMA connector, this transmission line consists of a Cu contact. Three silver (Ag) wires connect the Cu to the loop antenna and transmission line on top of the SiC substrate, which are both formed from a $\break$ 100-nm-thick Au contact layer with an approximately 5-nm layer of Ti to allow the Au to stick to the SiC substrate. The SiC substrate with the loop antenna and transmission line are contained within an Al holder, which possesses a Cu base. The dimensions of this holder are shown in Fig. \ref{fig: SensorHeadSiCAntenna}, alongside the microwave antenna and transmission-line design. The holder acts as a Faraday cage, ensuring that the microwaves are kept within the sensor head. The thickness of this holder sets the minimum sample to diamond distance at approximately 2 mm. 

In our previous work, an Al substrate with a copper transmission line and antenna loop, was selected due to its high thermal conductivity, which helps to mitigate temperature fluctuations in the diamond that produce noise that looks like a magnetic field \cite{acosta2010temperature}. Additionally, in contrast to FR4, Al does not burn under high laser powers. However, the microwave transmission properties of Al are inferior to those of FR4 and thus a material that combines good microwave and thermal properties is desired. SiC has a higher thermal conductivity than Al, 490 W/mK as opposed to 251 W/mK (at best), 
while the dielectric nature of SiC also ensures the free passage of microwaves, in contrast to metallic Al. 
This allows us to employ less microwave power and also obtain superior contrasts. Further improvement in microwave-power delivery efficiency is likely possible, via the use of superior SMA cables and solder. 


\section*{Appendix D: Parameter Optimisation}

For parameter optimization ODMR spectra are taken with the bias field aligned along a [100] orientation, as seen in Fig. \ref{fig: ODMRExample}. The ZCS, determined by applying a linear fit to the derivative slope of the higher-frequency NVC resonance of the demodulated ODMR spectrum, directly relates the LIA voltage output with a change in fluorescence output induced by a resonance-frequency shift. Larger ZCS values indicate greater responsivity and thus sensitivity; however, the ZCS does not account for greater susceptibility to noise and thus a larger ZCS does not always imply superior sensitivity.

\begin{figure}[h!]
\includegraphics[width=\columnwidth]{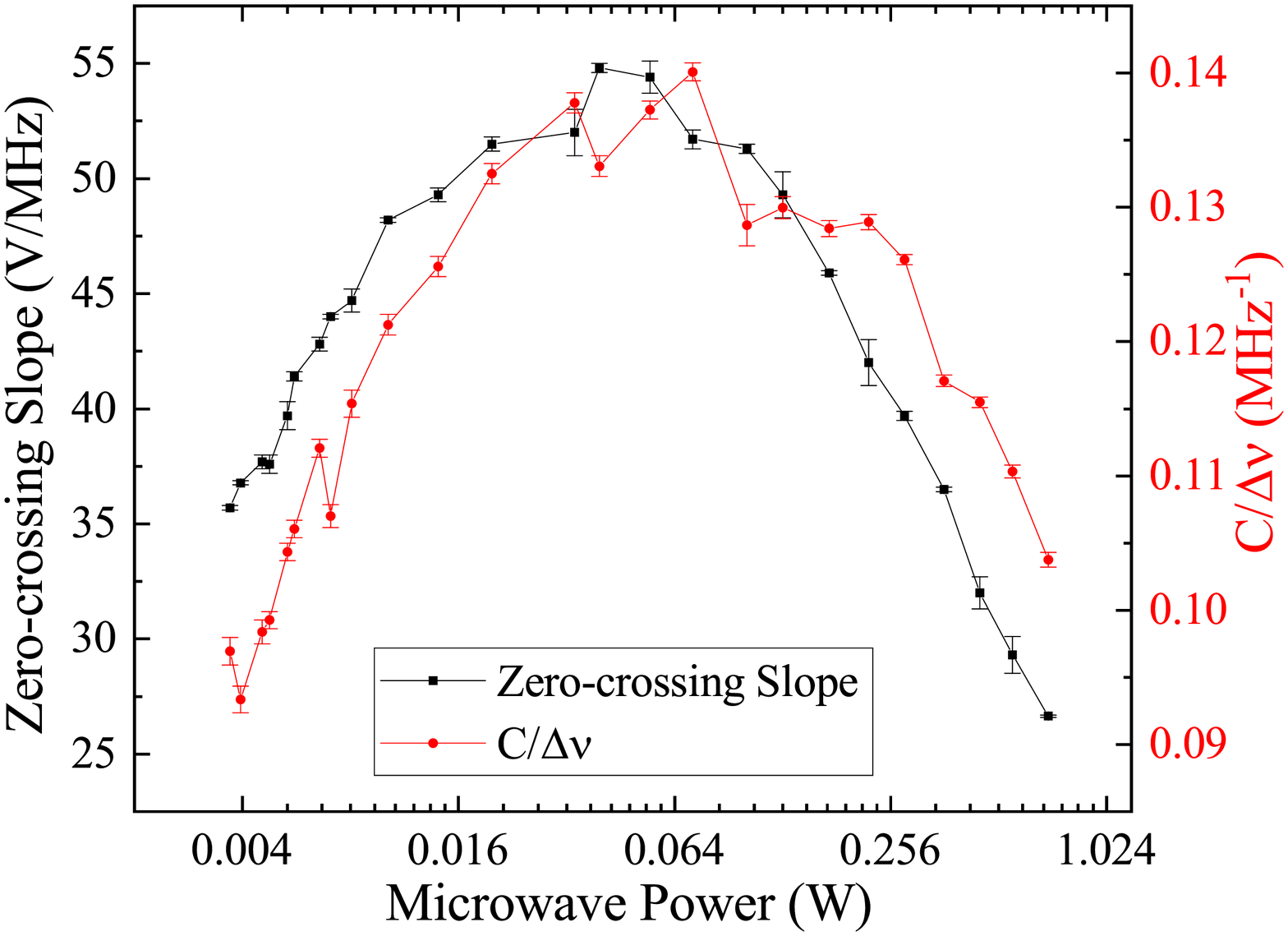} 
\caption{\small The ZCS and C/$\Delta$$\nu$ ratio as a function of the microwave power. All measurements are taken using a modulation frequency of 3.003 kHz and a modulation amplitude of 220.23 kHz.} 
\label{fig: Linewidth/ContrastvsMWPower}
\end{figure}

From Eq. (\ref{eq:PhotonShotNoiseLimit}), it can be seen that increasing the C/$\Delta$$\nu$ ratio improves the photon-shot-noise-limited sensitivity. Figure \ref{fig: Linewidth/ContrastvsMWPower} shows both C/$\Delta\nu$ and the ZCS as a function of the microwave power. For a given set of parameters, ODMR spectra are taken with and without the LIA, using hyperfine excitation. The contrast and line width can be extracted from applying Lorentzian fits to the pre-LIA ODMR spectra, taking the fluorescence signal directly from the signal photodiode. The ODMR spectra are taken with a frequency-sweep range of 2.88 - 2.92 GHz, a step resolution of 20 kHz, a step dwell time of 3 ms, and a sampling rate of 20 kHz (2 megasamples over $\break$ 100 s). The LIA 3-dB point is set to 500 Hz, with a filter slope of 48 dB per octave, and the LIA output scaling is set to 100. The reference input phase of the LIA is adjusted to maximize the signal within the $\textit{X}$-channel of the 
LIA. The contrast is determined using  

\begin{equation}
\label{eq:Contrast}
C = \frac{I_{O} - I_{R}}{I_{O}},
\end{equation}

where $I_{R}$ is the fluorescence signal on-resonance and $I_{O}$ is the fluorescence signal off-resonance. The line width is taken from the full width at half maximum (FWHM) of the Lorentzian fit. 
The optimum microwave power is found to occur when we observe approximately 0.054 W using an Agilent N9320B vector network analyzer. 
The stated microwave intensity, in watts, is the total power being delivered when exciting all three hyperfine features simultaneously, corresponding to the central peak of the five hyperfine features, the ODMR peak for which the ZCS is determined. For each hyperfine line, $\break$ approximately 0.018 W of microwave power is being delivered. That this is the optimum value can also be seen from the decrease in ZCS beyond this microwave power. 
The microwave powers stated above neglect losses following the circulator within the cables, adaptors and transmission line and thus are very approximate and unlikely to reflect the true microwave power being delivered to the diamond sample. 

\begin{figure}[h!]
\includegraphics[width=80mm]{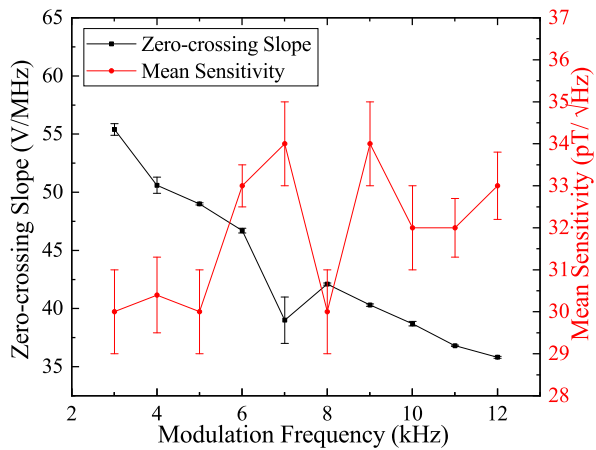} 
\caption{\small The ZCS and mean sensitivity, taken in a $\break$ (10 - 500)-Hz frequency range, as a function of the modulation  frequency. All measurements are taken using a modulation amplitude of 220.23 kHz and a microwave power of 0.054 W.} 
\label{fig: MeanSensvsFMRate}
\end{figure}

The ZCS is found to increase with decreasing modulation frequency as can be seen in Fig. \ref{fig: MeanSensvsFMRate}. This increase is expected due to the finite repolarization time of the NVC \cite{barry2016optical, el2017optimised, schoenfeld2011real}. However, the noise floors also rise, as susceptibility to noise is increased for modulation frequencies that approach dc.

Accordingly, we measure the mean sensitivity taken over a (10 - 500)-Hz frequency range as a function of the modulation frequency, as seen in Fig. \ref{fig: MeanSensvsFMRate}.  
A modulation frequency of 3.003 kHz is found to provide the best mean sensitivity over the given frequency range. 
For the vacuum-pump measurements care is also taken to ensure that for a given sampling rate, the modulation frequency does not lead to aliased peaks at,e.g., 1 kHz, the frequency of the turbo-pump signal. 

\begin{figure}[h!]
\includegraphics[width=\columnwidth]{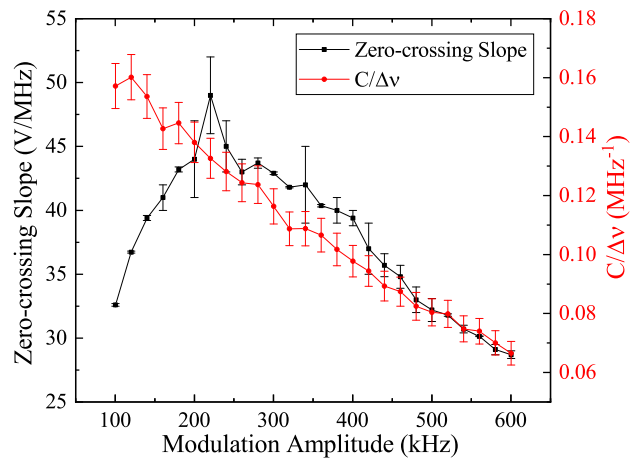} 
\caption{\small The ZCS and C/$\Delta$$\nu$ ratio as a function of the modulation amplitude. All measurements are taken using a modulation frequency of 3.003 kHz and a microwave power of 0.054 W.} 
\label{fig: ZCSvsFMDepth}
\end{figure}

Figure \ref{fig: ZCSvsFMDepth} shows the ZCS and C/$\Delta$$\nu$ ratio as a function of the modulation amplitude. A modulation amplitude of 220.23 kHz is found to provide the highest ZCS, with no change being observed in the noise floor over the range of modulation-amplitude values. It would appear that further optimization of the modulation amplitude may be possible, however, as when frequency modulation is used, a line width of 0.73 MHz is measured, compared to a line width of 0.63 MHz when it is off. The C/$\Delta$$\nu$ ratio continues to increase below the modulation frequency for which the highest ZCS is measured: this suggests that the line-width broadening caused by the frequency modulation is not the dominant effect influencing the ZCS below 220.23 kHz
. From theoretical considerations, it is expected that the optimum modulation amplitude will be equal to $\Delta$$\nu$/2$\sqrt{3}$, which for a line width of 0.63 MHz would mean a modulation amplitude of approximately 180 kHz \cite{barry2016optical}. 

\begin{figure}[h!]
\includegraphics[width=\columnwidth]{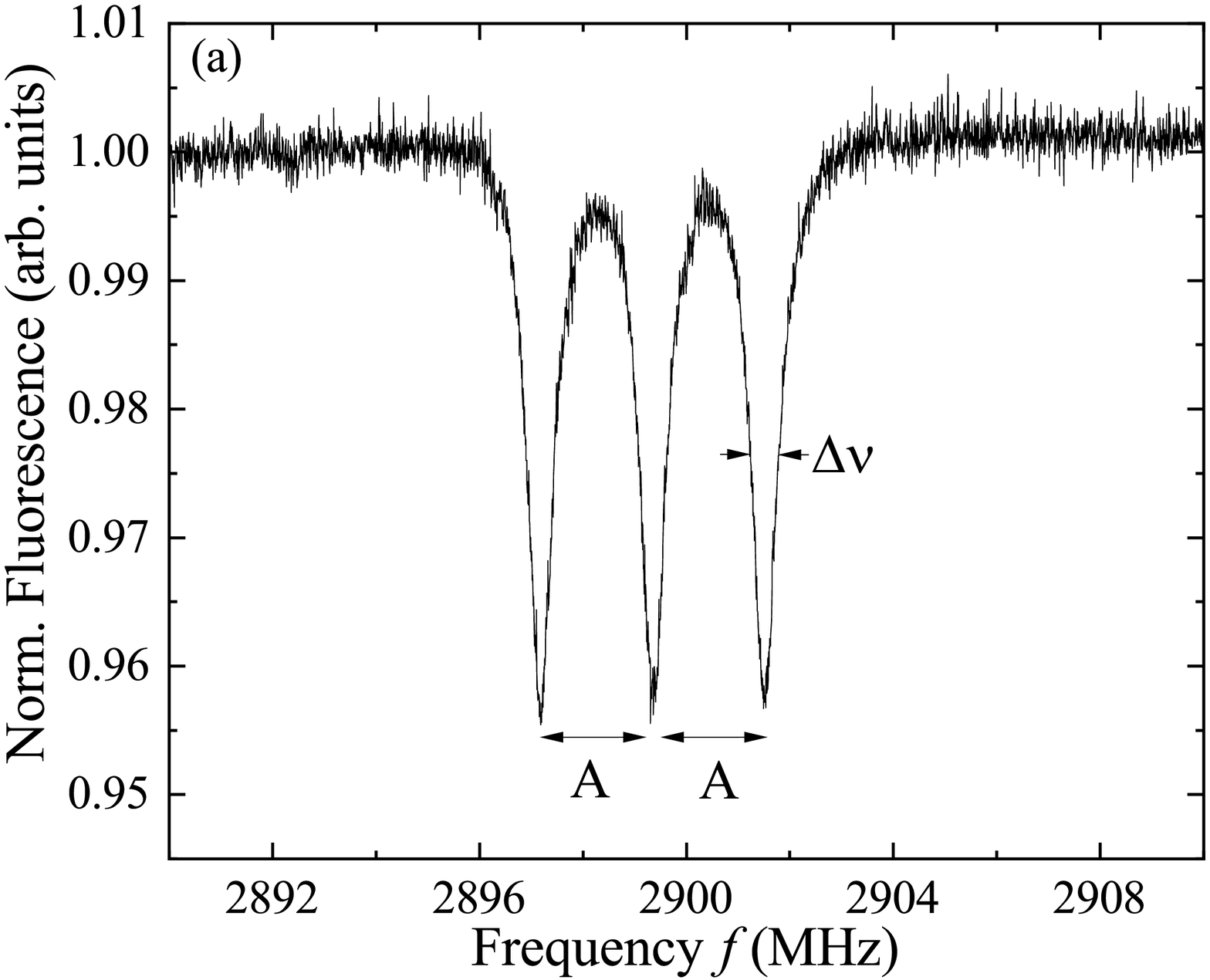} 
\includegraphics[width=\columnwidth]{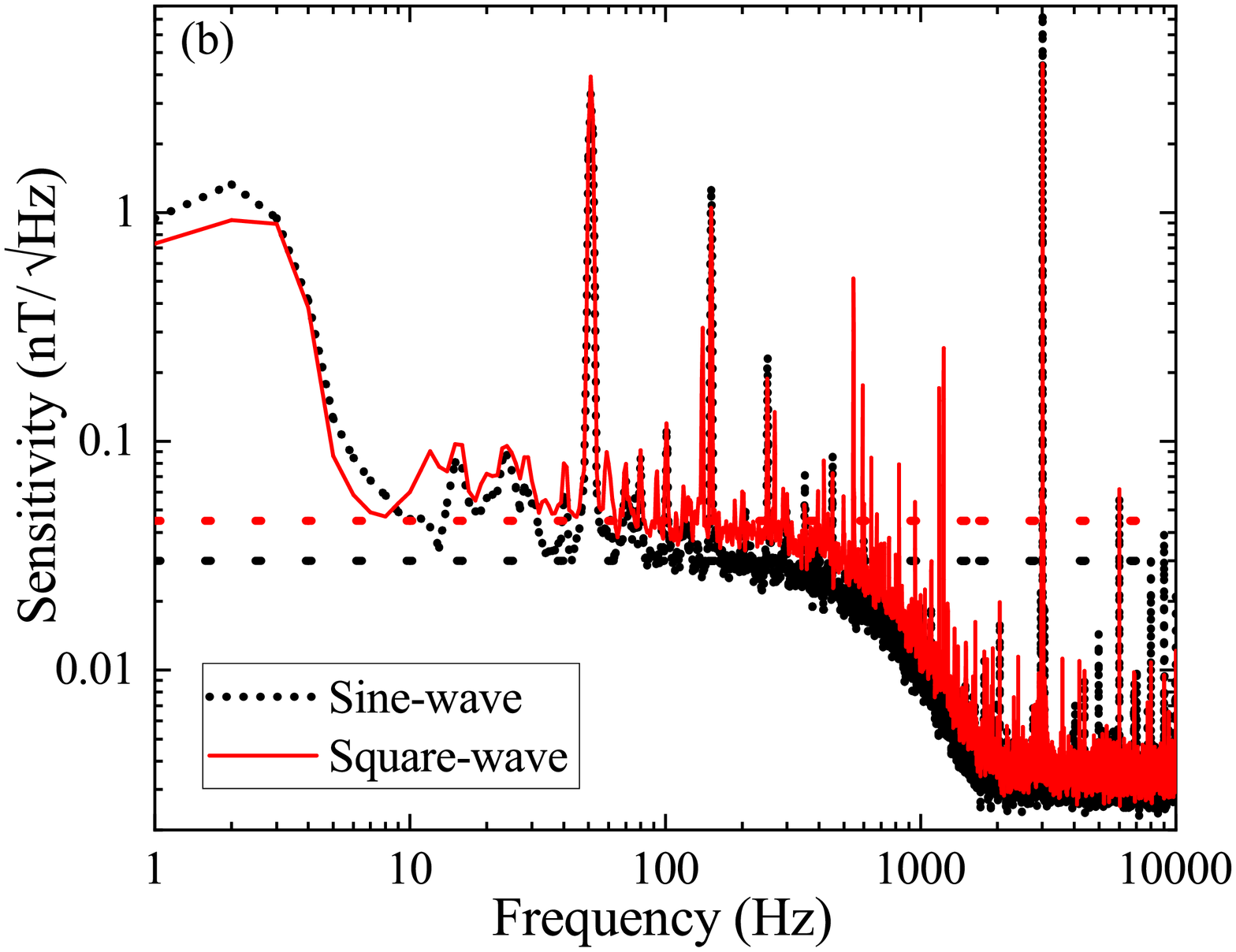} 
\caption{\small (a) The ODMR spectrum, showing the hyperfine splitting, \textit{A} and the line width $\Delta$$\nu$. Hyperfine excitation is not being used. b) Sensitivity measurements showing the difference in sensitivity between sine and square-wave frequency modulation using the optimum settings. The dotted lines show the mean sensitivity taken for a frequency range from 10 to 500 Hz, with the 50-Hz peak and its harmonics masked. Mean sensitivities of (30 $\pm$ 10) pT/$\sqrt{\textrm{Hz}}$ and $\break$ (45 $\pm$ 20) pT/$\sqrt{\textrm{Hz}}$ are found for sine and square-wave frequency modulation, respectively: ac coupling is used for these measurements.} 
\label{fig: SinevsSquare}
\end{figure}

In this work we use sine as opposed to square-wave frequency modulation. Figure \ref{fig: SinevsSquare}(b) shows sensitivity measurements taken with sine and square-wave modulation. The noise floor is found to be significantly higher for square-wave modulation: this is likely due to the generation of multiple additional harmonics when using square-wave modulation. A negligible difference in the ZCS is seen between sine and square-wave modulation, this is surprising given that our line width ($\Delta$$\nu$ = 0.73 MHz) to hyperfine splitting (A = 2.158 MHz) ratio is greater than 1/4, although we are only using a modulation amplitude of 220.23 kHz \cite{el2017optimised}. This may be a property of the E8257 microwave source and is not observed using the $\break$ Agilent N5172B microwave source \cite{patel2020subnanotesla}. Figure \ref{fig: SinevsSquare}(a) shows the hyperfine splitting and line width, without hyperfine excitation.

From this parameter optimization process the optimum modulation frequency, modulation amplitude, and microwave power are identified to be 3.003 kHz, $\break$ 220.23 kHz, and 0.054 W respectively. A laser power of 1.8 W, corresponding to an excitation power at the signal-sensor-head diamond of approximately 0.3 W, is used. Continuing to increase the laser power beyond this causes the photodiodes to first enter a nonlinear response regime and eventually to saturate. It is possible that given photodiodes with higher dynamic range, the use of higher laser powers could yield improvements in sensitivity, with increases to the photon detection rate, $\textit{R}$. However, increasing the laser excitation power will also change the spin polarization rate, cause power broadening, and lead to increased noise from temperature fluctuations. 
Accordingly, it would be more desirable to improve the photon detection rate via enhancements in the photon collection efficiency.

\FloatBarrier

\section*{Appendix E: Optics}

The green-to-red photon conversion efficiency is $\break$ approximately 0.36\%, enabling a red-photon detection rate of $\break$ 2.9 $\times$ $10^{15}$ Hz for approximately 0.3 W of laser excitation power. 
This is achieved via a series of incremental improvements in both the sensor-head and table-top optics. First, the replacement of a 400-$\mu$m-diameter 5-m long 0.22-NA FG400AEA fiber with a 910-$\mu$m-diameter 3-m long 0.22-NA FG910UEC fiber is found to improve the observed fluorescence signal for a given laser power. It is easier to couple both the laser beam and fluorescence into the larger-diameter fiber and the reduced length allows them to be better secured to minimize modal noise. Larger-diameter fibers are also anticipated to be less susceptible to modal noise due to them possessing a higher number of modes and thus a reduced speckle contrast \cite{epworth1978phenomenon}. Using a Thorlabs CS165CU/M $Zelux^{TM}$ color CMOS camera the standard deviation of the fiber-output speckle pattern is measured to be 0.42 and 0.39 for the FG400AEA and FG910UEC fibers respectively.  
However, the 910-$\mu$m fiber is more rigid and we are unable to procure a ceramic FC-PC termination for the sensor head end of the fiber. The first of these factors would be inconvenient if scanning the sensor-head across a sample, while the use of a steel FC-PC termination could potentially act as a source of magnetic noise if the steel moves by even a small amount relative to the diamond. Introducing padding should help to reduce the effects of vibrations on the sensor head and help with 1/$\textit{f}$ noise. Given our narrow line widths and [100] bias-field alignment, relative motion-e.g., from vibration - between the sensor head and the permanent $\break$ Nd-Fe-B magnet is likely to be a not insignificant source of low-frequency noise. The excitation and collection efficiency could also be improved by better securing of the optical fibers; movements and misalignment of the fibers are found to cause large drops in fluorescence. This not only contributes to noise, not readily cancelled using the balanced detector, but also leads to inefficient collection and excitation, as the skew in the light launched into the fiber produces a "donut-shaped" beam output \cite{wang2010propagation}. 
Some improvement in fluorescence collection may also be the result of the highly reflective SiC substrate upon which the diamond is placed. 

As discussed in Appendix C the sensor-head optics are contained within a 25.4 mm-diameter SM1 open lens tube that is 20 mm in length. 
This arrangement allows us to observe what happens to the fluorescence emitted by the diamond. We find that it rapidly diverges before it can be effectively collimated and focused by the lenses. The use of a single lens tube allows us to then bring the lenses closer together and to control the relative positions of the fiber, C220TMD-B, and C330TMD-B lenses in the z-direction. Using a Thorlabs CXY1 $\textit{XY}$ translating lens mount it is possible to obtain $\textit{x}$ and $\textit{y}$ control in addition to z-control, although we found that the fiber, lenses, and diamond are already well-centered without this additional degree of freedom. 
The 0.25 NA C220TMD-B ashperic lens is used both for collimation in the sensor head and for coupling the laser beam into the fiber on the table-top optics side. This NA closely matches that of the 0.22-NA fiber. 

For the table-top optics 
Thorlabs BB111-E02 $\break$ $>$ 99\% reflectance Zerodur  broadband dielectric mirrors are used, helping to minimize laser and fluorescence losses. 
Thorlabs DMSP605 short-pass dichroic mirrors, which possess a 99\% reflectance at 650 nm, are used to separate the green laser and red fluorescence light. 
Imaging with the CS165CU/M $Zelux^{TM}$ color CMOS camera allows us to ensure that the fluorecence beam diameter matches the 0.8-mm diameter of the balanced detector photodiodes. 
15-mm diameter, 12-mm-focal-length, Thorlabs ACL1512U aspheric condenser lenses are used, in combination with 25-mm-focal-length aspheric lenses, to achieve this beam diameter. A tighter focus onto the photodiodes would be undesirable, as it could lead to nonlinear effects. 

The use of two sensor heads necessitates the use of more laser power than for a single-sensor-head configuration. The 1.8 W of laser power is split approximately equally between the two sensor heads using a Thorlabs PBS12-532-HP high-power polarizing beam splitter (PBS) and a Thorlabs WPH05M-532 zero-order half-wave plate, held within a Thorlabs RSP1/M rotation mount. When on a magnetically sensitive resonance and using a Thorlabs PM100D power meter with S121C head the laser power prior to the fibers is found to be $\break$ approximately 0.6 W and 0.75 W for the signal and reference sensor heads respectively. Given a measured approximately 80\% transmission efficiency through the optical fiber, and further losses from the sensor-head lenses, in part due to their B (650 - 1050 nm) antireflection coatings, which are selected to maximize fluorescence even at the expense of green laser power, $\break$ approximately 0.3 W of laser excitation power is directed upon the diamond. For the reference sensor head, this number is $\break$ approximately 0.4 W. On the magnetically sensitive resonance, $\break$ approximately 0.9 mW of fluorescence is measured just prior to the signal and reference photodiodes, suggesting that the optics alignment for the reference channel could be improved. Considering only the laser power going to the signal sensor head, the green-to-red photon conversion efficiency is found to be approximately 0.36\% or 0.18\% for whether or not the laser-power losses prior to the diamond are included, respectively. A lower total laser power could be employed via the use of optics with AB (400 - 1100 nm), as opposed to B antireflection coatings. Additionally, the optical isolator is found to be responsible for approximately 0.2 W of the laser-power loss. The coupling of the laser-beam output of the fibers into the sensor heads could also likely be improved. Small changes in the positioning of the fiber are found to yield large changes in the laser power measured at the diamond as well as in the fluorescence signal measured at the photodiodes. If the fiber is moving around during measurements changes in the coupling efficiency could be a substantial source of noise, alongside modal noise.

\section*{Appendix F: Sensitivity Data Analysis}

\begin{figure}[h!]
\includegraphics[width=\columnwidth]{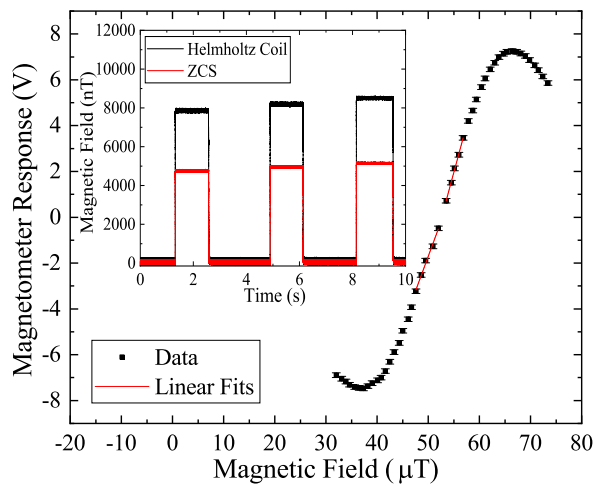}
\caption{\small The magnetometer response as a function of the applied magnetic field along [100], with the polarity of the Helmholtz coil reversed to observe the symmetry of the magnetometer response for both directions (along and against the bias field) of the given [100] orientation. Linear fits are made on the linear section for each polarity of the Helmholtz coil. The inset shows some of the step signals observed when applying the test fields. The voltage shift is converted into magnetic field units using either the test-field (black) or ZCS (red) calibration constant, showing the effect of the projection along the NVC axes on the measured field.} 
\label{fig: StepSignal-TestFieldMeasurement}
\end{figure}

In this work, to determine the sensitivity 1-s time traces are taken using a PicoScope 5442D at a sampling rate of 20 kHz and these are converted into PSD plots via MATLAB. Care is taken to ensure that time traces and their corresponding MATLAB PSD plots agree with Parseval's theorem, 

\begin{equation}
\label{eq:ParsevalTheorem}
E = \int_{-\infty}^{\infty} x(t)x^{*}(t)\,dt\ = \frac{1}{2\pi}\int_{-\infty}^{\infty} X(\omega)X^{*}(\omega)\,d\omega\,
\end{equation}

that the energy of the signal should be conserved between the time and frequency domain \cite{hobson2006mathematical}. In Eq. (\ref{eq:ParsevalTheorem}) E is the energy of the signal, \textit{x(t)} is the signal in the time domain and $X(\omega)$ is the signal in the frequency domain. The time traces in volts (V) are converted into nanotesla (nT) by dividing through by a calibration constant with units of V/nT. This calibration constant is found via either the test-field or ZCS method. In the test-field approach a field from a Helmholtz coil, as explained in the main text, is applied along [100] while sitting on a magnetically sensitive resonance. The shift in voltage due to the applied field is measured as the current through the Helmholtz coil is increased in steps. The Helmholtz coil has been calibrated using a Magnetics GM07 Hall probe that measures the total field. Fluctuations in temperature and laser power can also cause voltage shifts and thus the magnetic field is turned off between each increase in test field magnitude, allowing potential changes in the zero-test-field voltage to be taken into account. Examples of the step signals observed when applying the test-field can be seen in the inset of Fig. \ref{fig: StepSignal-TestFieldMeasurement}. The calibration constant in V/nT can then be found by applying a linear fit to the data seen in Fig. \ref{fig: FFT-Sensitivity} of the main text. The alternative, ZCS method involves applying a linear fit to the zero-crossing point of the ODMR resonance to obtain a slope in V/MHz. This slope can then be converted into an appropriate calibration constant using the gyromagnetic ratio of the NVC, $\break$ $\gamma$ = 28 Hz/nT. A ratio of approximately 0.58 is found between the calibration constants determined using the test-field and ZCS methods respectively. This is due to the projective nature of NVC magnetometers, with the actual field measured along each of the four NVC alignments being a factor of 0.58 smaller than the total field measured with the Hall probe. The ZCS method is not employed for this work, as with the bias field aligned along [100] the sensitive axis is now along [100] as opposed to a $\langle$111$\rangle$ NVC symmetry axis. The calibration constant determined using this method would be as if the total magnetic field had been completely projected along each of the four NVC alignments symmetry axes simultaneously: a situation that would never occur in practice. 

FFTs in decibel units (dBu), referenced to a voltage of 0.775 V, are also taken of the 1-s time traces using the PicoScope 5442D. Then these FFTs are extracted and converted into V using the conversion formula $\break$ 0.775 $\times$ $10^{L/20}$ where $\textit{L}$ is the FFT noise floor in dBu \cite{patel2020subnanotesla}. The calibration constant can then be used to convert these FFTs into units of nT/$\sqrt{\textrm{Hz}}$. For the FFTs taken both in PicoScope and directly from the time traces in MATLAB a Blackman window is used, with the appropriate correction factors being taken into account. The sensitivities determined via MATLAB and using the PicoScope FFTs are found to be in agreement. When determining the mean sensitivity the 50 Hz peak and its harmonics at 100 Hz, 150 Hz, $\break$ 200 Hz, 250 Hz, 300 Hz, 350 Hz, 400 Hz, and 450 Hz are treated as signals, given their known origin, and masked. Others peaks between 10 - 500 Hz are not masked. The standard deviation of the mean of the sensitivity spectrum in a 10 - 500 Hz frequency range is taken as the error. It should be noted that for all our measurements the higher frequency ODMR resonance is used, as this is the resonance for which the microwave power has been optimised. The difference in ODMR contrast and linewidth between the two ODMR resonances that can be observed in Fig. \ref{fig: ODMRExample}(a) and Fig. \ref{fig: ODMRExample}(b) in the main text are a consequence of the amount of microwave power differing between the two frequencies, as is observed using an Agilent N9320B vector network analyser. Sensitivity measurements are taken using the same LIA and microwave settings as their corresponding ODMR spectra, with a 100 LIA output scaling.

\section*{Appendix G: Balanced Detection}

\begin{figure}[h!]
\includegraphics[width=\columnwidth]{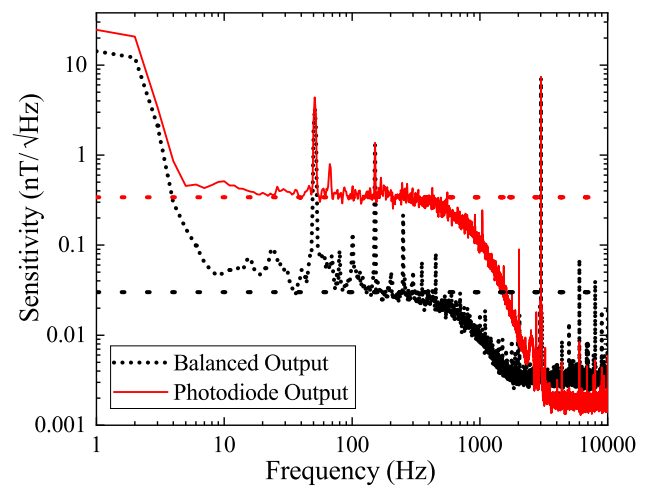} 
\caption{\small Sensitivity measurements taken with and without the use of "red-red" balanced detection. Mean sensitivities of (30 $\pm$ 10) pT/$\sqrt{\textrm{Hz}}$ and (340 $\pm$ 50) pT/$\sqrt{\textrm{Hz}}$ are found with and without balanced detection, respectively. No ac coupling is used for these measurements.} 
\label{fig: BalancedDetectionFFT}
\end{figure}

In this work a second reference sensor head containing a diamond is employed to provide fluorescence for balanced detection as opposed to picking off a portion of the laser beam; in other words, "red-red" noise cancellation. 
As discussed in appendix B this is also a $\break$ 1 $\textrm{mm}^3$ diamond cut from the same original sample. It is important for the two sensor heads to be as identical to one another as possible. The optics arrangement used for the signal and reference paths are very similar, though it is clear there is some difference in the excitation power and collection efficiency for each of the sensor heads. When on a magnetically sensitive resonance the rotation mount containing the wave plate is adjusted until the balanced output is set to zero, at which point the fluorescence signals from each sensor head are equal. 
When off resonance the fluorescence level from the signal sensor head is higher than that from the reference sensor head. 
Figure \ref{fig: BalancedDetectionFFT} shows the effect of using the balanced output as
opposed to the main signal output without any noise cancellation. "Red-red" noise cancellation is found to provide a factor-of-11 improvement in sensitivity. Without common-mode noise cancellation the sensitivity is limited by laser noise and also other sources of noise common to each sensor head. 
Nonetheless, this does not appear to be notably superior to a scheme employing picked off green laser light and, as the sensitivity measurements in Fig. \ref{fig: FFT-Sensitivity} show laser noise remains a significant part of our overall noise floor \cite{patel2020subnanotesla}. Further improvement in this "red-red" noise cancellation scheme is likely possible via attempts to equalize the laser power being delivered to each sensor head, as increased laser excitation power in one sensor head will, e.g., lead to differing levels of temperature fluctuation and thus noise. Additionally, noise associated with the optical fibers likely differs for each of the two sensor heads. With these further optimizations it may yield superior noise cancellation to a picked-off green scheme. 

Gradiometry or differential magnetometry would allow magnetic noise to be cancelled, assuming that it is common to each sensor head \cite{blakley2015room, blakley2016fiber, masuyama2021gradiometer, zhang2021diamond}. Effective cancellation of magnetic noise could enable us to get down to the magnetically insensitive noise floor of $\break$ (21 $\pm$ 4) pT/$\sqrt{\textrm{Hz}}$. Gradiometry could be implemented using the current setup by supplying microwaves to the reference sensor head. The reference sensor head, at $\break$ approximately 180 $^{\circ}$ to the signal sensor head, is presently aligned such that the bias field is also along a [100] orientation and the sensor heads show close to identical performance, which is necessary for the effective cancellation of magnetic noise.  
It would be convenient to use two separate microwave sources to address each sensor head; however, if a single microwave source could be employed alongside a microwave splitter, it would also allow for the common microwave phase noise to be cancelled, although this does not appear to be a limiting source of noise at present. 

\FloatBarrier

\section*{Appendix H: Power Saturation}

\begin{figure}[h!]
\includegraphics[width=\columnwidth]{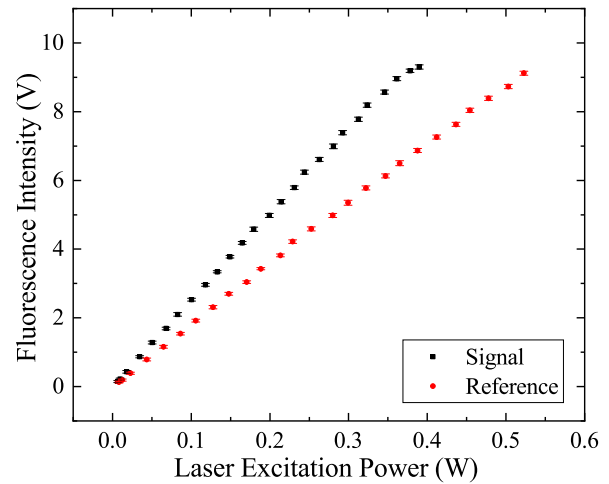} 
\caption{\small Power-saturation curves showing the fluorescence intensity at the photodiodes as a function of the laser excitation power for both signal and reference sensor heads. } 
\label{fig: PowerSaturationMeasurement}
\end{figure}

Figure \ref{fig: PowerSaturationMeasurement} shows power-saturation measurements for both the signal and reference sensor heads. It is clear that we are far from saturating the NVC ensembles in either sensor head. Indeed, given the dynamic range of the photodiodes $\break$ (a 10-V maximum output voltage) we would not be able to saturate the NVC ensemble before the photodiodes. The saturation that appears in the signal-sensor-head saturation data is likely due to the response of the photodiode. This can be seen from the fact that the reference sensor-head data do not display the same saturation behavior, even with the higher laser excitation power. The signal and reference sensor heads differ in fluorescence intensity in part due to the fact that these are the off-resonance fluorescence intensities and thus the photodiode signals are unbalanced. It is also apparent that the green-to-red photon conversion efficiencies differ between the two sensor heads (approximately 0.27\% for the reference as opposed to approximately 0.36\% for the signal sensor head) from power-meter measurements of the red fluorescence taken when on a magnetically sensitive resonance. Accordingly, more laser power is required to obtain the same fluorescence intensity for the reference sensor head.

It is probable that the NVCs contained within the sensing volume directly illuminated by the focused laser beam are fully saturated at a excitation laser power of $\break$ approximately 0.3 W. Saturation is not observed from the power-saturation measurements. This is likely because as the laser excitation power is increased more photons are scattered through the diamond and thus excite fluorescence from NVCs outside of the immediate sensing volume. This means that the sensing volume has been increased greatly, perhaps to encompass the entire diamond sample, which could be negatively impacting performance depending upon the microwave and bias-field homogeneity over the entire diamond sample. Increasing the laser power in an effort to saturate the NVC ensemble will exacerbate these homogeneity issues.

\section*{Appendix I: Hyperfine Excitation Scheme}

\begin{figure}[h!]
\includegraphics[width=\columnwidth]{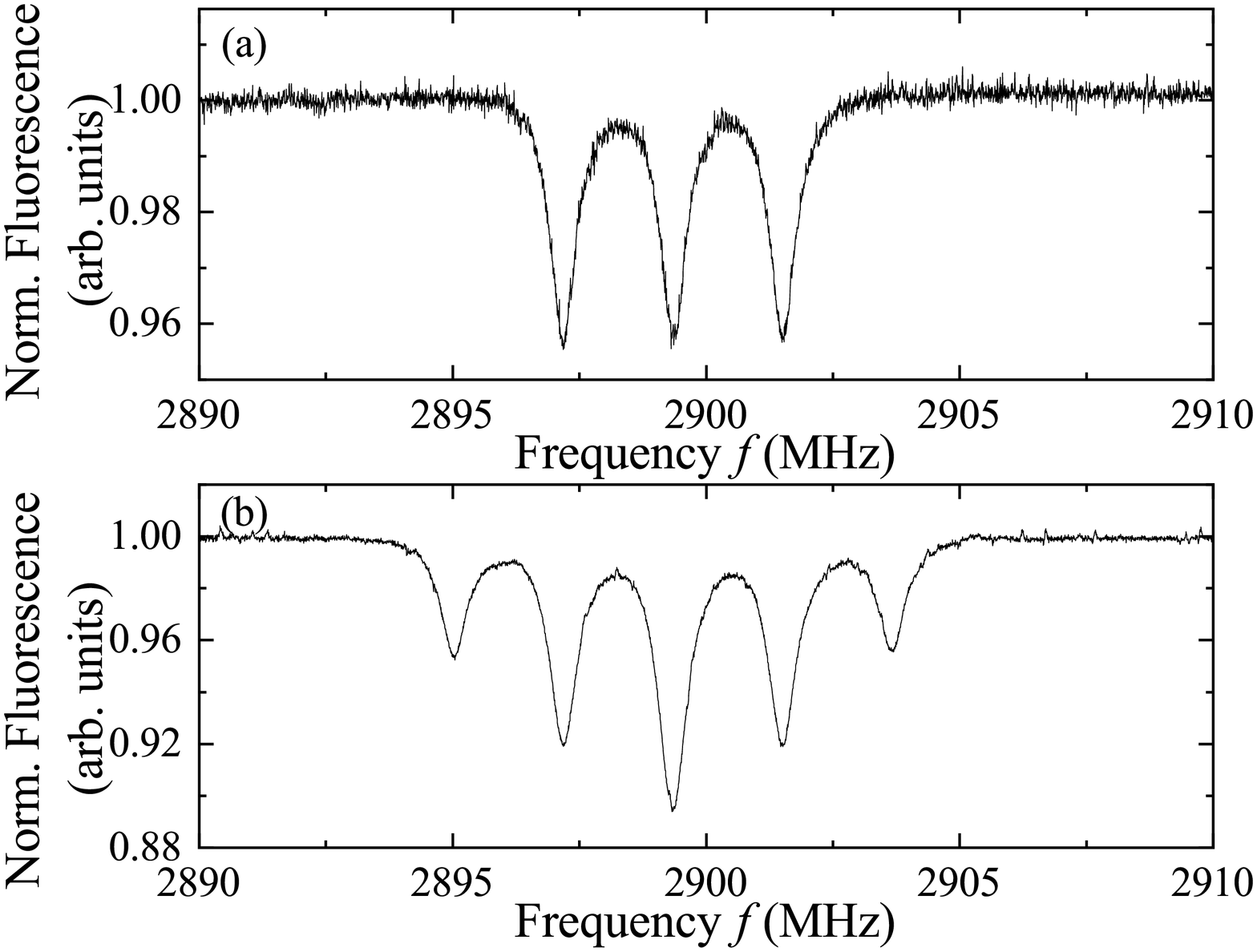} 
\includegraphics[width=\columnwidth]{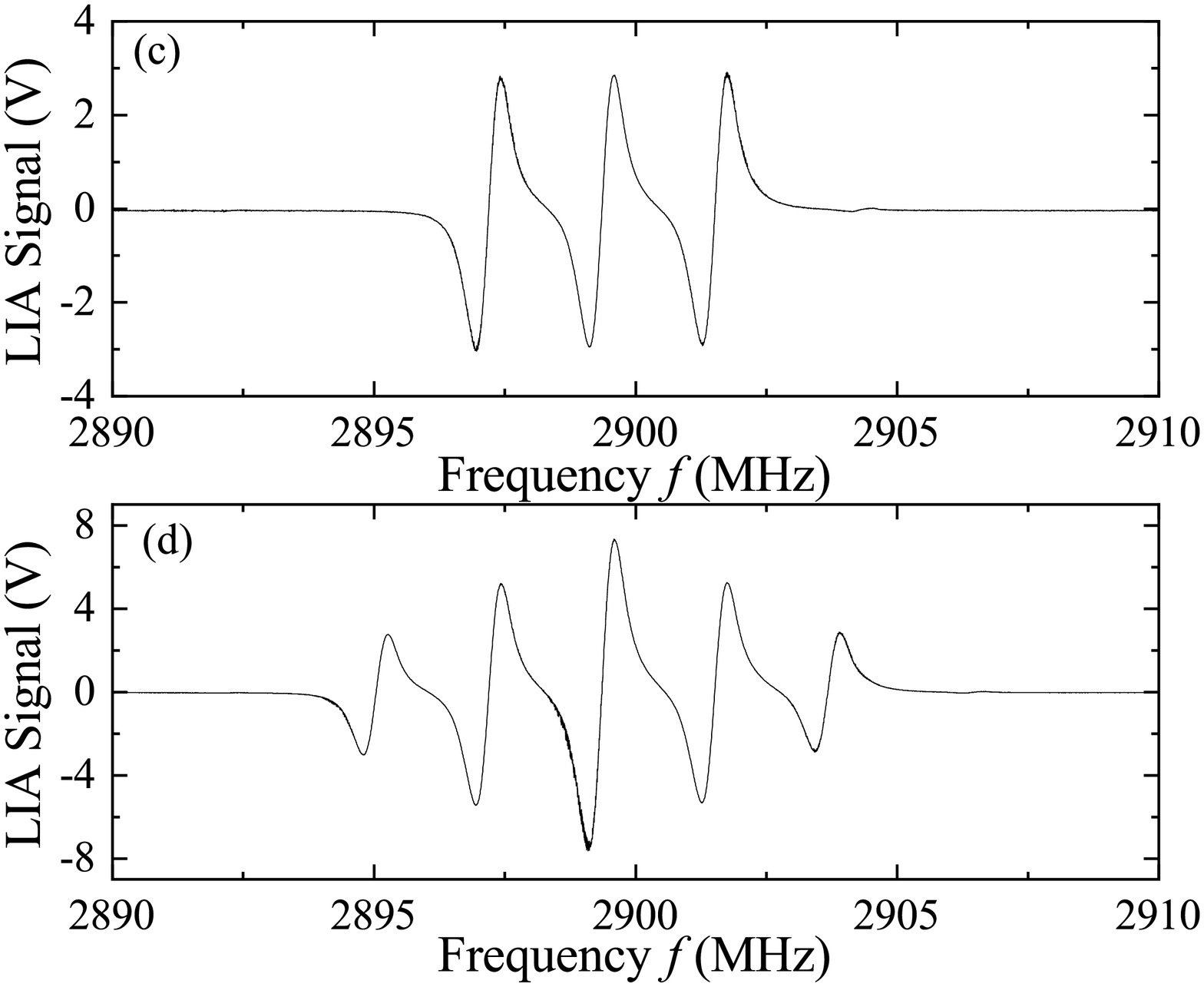}
\caption{\small (a),(b) The ODMR spectra (a) with and (b) without hyperfine excitation. (c),(d) The demodulated ODMR spectra (c) with and (d) without hyperfine excitation.} 
\label{fig: HFOnvsOff}
\end{figure}

Figure \ref{fig: HFOnvsOff} shows the ODMR spectrum taken, with the bias field aligned along one of the [100] orientations, with and without hyperfine excitation. Hyperfine excitation is found to improve the contrast from 4\% to 10\%, a factor-of-2.5 improvement. 
However, the line width is also broadened from 0.70 MHz to 0.73 MHz due to increased microwave-power broadening. 
Further parameter optimization may be possible. 
We use an external Mini-Circuits ZX05-U432H-S+ up-converter, 
and the output of the microwave source and the RSPro AFG21005 arbitrary function generator are sent into the local-oscillator and intermediate-frequency inputs of the up-converter respectively. The RF output is then sent to a 43-dB Mini-Circuits  ZHL-16W-43-S+ microwave amplifier, followed by a Pasternack PE83CR005 (2-4)-GHz circulator, and thence onto the sensor head.

\begin{figure}[h!]
\includegraphics[width=\columnwidth]{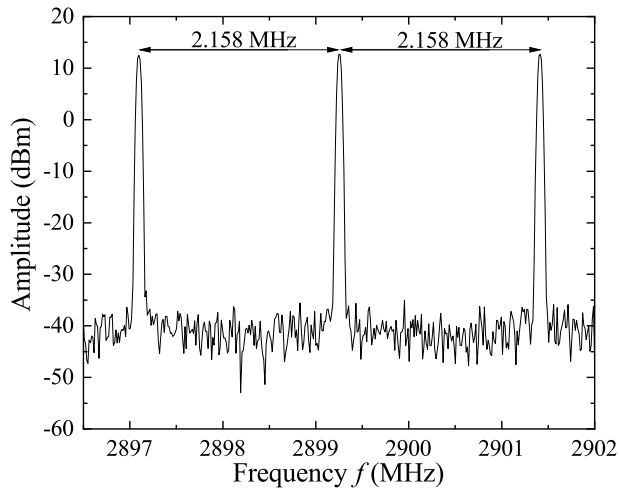} 
\caption{\small The three microwave tones used for hyperfine excitation as observed using a vector analyzer. The Bayonet Neill-Concelman (BNC) cable used to connect the subminiature type A (SMA) cable to the vector analyzer is found to cause approximately 8.5 dB in microwave losses. These losses are accounted for when determining the microwave intensity of each tone.} 
\label{fig: MWTone}
\end{figure}

The three microwave tones can be observed using an Agilent N9320B vector network analyzer. It is thus possible to vary the amplitude of the sine-wave and microwave power to identify combinations that yield three microwave tones of approximately equal intensity, an example of which can be seen in Fig. \ref{fig: MWTone}. 
This corresponds to a total microwave power of approximately 0.054 W on the vector analyzer for the centermost hyperfine feature, distributed across the three microwave tones each of which has $\break$ approximately 0.018 W of power, which is identified to be the optimum microwave power via the parameter optimization process described in appendix D. For the nonhyperfine case, a single microwave tone of approximately 0.018 W of microwave power is used. Increasing the microwave power to $\break$ approximately 0.054 W for a single microwave tone does not yield the contrast improvement seen from hyperfine excitation but does lead to considerable line-width power broadening. 
The ratio of the peaks observed in Fig. \ref{fig: HFOnvsOff} is 1:1.7:2.3:1.7:1, being taken relative to the first of the five hyperfine-excitation ODMR features. The contrast increase of 2.3 relative to the first peak when using hyperfine excitation as opposed to when comparing the contrast to single excitation suggests a discrepancy in the microwave intensity of a single tone between the two sets of measurements in Fig. \ref{fig: HFOnvsOff}. These results are in closer agreement to theoretical predictions than our previous work and they enable a further improvement in contrast and thus sensitivity \cite{barry2016optical, patel2020subnanotesla}. Nonetheless, the line widths of each of the five ODMR peaks still differ from around 0.66 MHz for the single-tone outer peaks to 0.73 MHz for the central peak corresponding to three-tone excitation. It is apparent 
that there remains a not insignificant difference in the microwave intensity of each tone. Further improvement in the hyperfine-excitation scheme should be possible given additional optimization of the power balance between the microwave source and the function generator. Additionally, the balance of the tones is observed via the vector analyzer and this may be misleading, as it would not account for microwave power losses and other effects that may occur on the path to the diamond. Working through the large parameter space of microwave power and sine-wave amplitude combinations would be required to optimize the performance further. 

\section*{Appendix J: Frequency Range}

The upper limit of the sensitive frequency range is taken to be the 3-dB point of the LIA LPF, which is set to 500 Hz with a filter order of 48 dB per octave. 
The lower limit is selected to be 10 Hz due to the 1/$\textit{f}$ noise, which appears to be associated primarily with environmental magnetic noise, below this point. A 3-dB point of up to 2.8 kHz is employed for the backing-pump and turbo-pump measurements, however, in order to ensure that the filter response does not cut off any of the magnetic signals from the machinery. When using this higher 3-dB point value, the 3.003-kHz modulation frequency comes to dominate the PicoScope time traces that are subsequently converted into PSDs. To continue collecting time traces via the PicoScope 5442D with this combination of 3-dB point and modulation frequency requires the use of a higher time-trace $\textit{y}$-axis (voltage) range, resulting in an inferior resolution. 
To allow a consistent voltage resolution to be maintained, the modulation frequency is thus set to 10.003 kHz. This does lead to a lower ZCS, as seen in Fig. \ref{fig: MeanSensvsFMRate}, limiting the sensitivity and hence the sensing range for the vacuum-pump measurements. The increased sensitivity at $\break$ 3.003 kHz, however, would not lead to an especially significant increase in sensing range given the 1/$d^2$ drop-off observed in Fig. \ref{fig: BackingPumpOnvsOff}(b). In principle, the maximum frequency range that can be obtained with this magnetometer would be limited by the modulation frequency and, ultimately, the response of the NVCs, in terms of the repolarization rate \cite{fescenko2020diamond, xie2020dissipative}. At present we are limited by the frequency response of the LIA LPF. The sampling rate would not limit our ability to detect signals in this case, given it is set to 20 kHz, and thus the Nyquist frequency is significantly higher than both our modulation frequency and the frequencies of the machinery signals of interest.

\section*{Appendix K: Dynamic Range}

One of the main benefits of NVC magnetometers is their potential for high dynamic range \cite{balasubramanian2008nanoscale}. The dynamic range of the NVC magnetometer ($\delta B$) is given by $\break$ $\delta B$ = $\delta$ $\textit{f}$/$\gamma$, where $\delta f$ is the line width of the NVC resonance. From this and using a line width of 0.73 MHz the dynamic range is determined to be $\pm$ 13.1 $\mu$T \cite{wang2022portable}. This agrees with results obtained when applying a test field along [100] using the Helmholtz coil. Figure \ref{fig: StepSignal-TestFieldMeasurement} shows the voltage shift as a function of the applied magnetic test field, with the polarity of the Helmholtz coil being switched to apply fields along both the +1 and -1 directions of a [100] orientation (along and against the bias field). A slight disparity is observed in the slope between the two polarities, suggesting that the test fields are not quite aligned along [100]. 
The test field is increased in magnitude until the resonance frequencies cease to lie on the central ODMR dispersion feature. 

Improvements in line width and thus sensitivity are at the expense of a decrease in dynamic range. This is problematic for the backing-pump measurements, because if the device is brought closer than approximately 20 cm, the dc magnetic field from the steel components of the pump begins to negatively impact the ODMR spectra in terms of contrast and line width, as well as causing voltage shifts out of the linear region that render the magnetometer magnetically insensitive. This could be accounted for by adjusting the bias-field alignment. More practically, however, feedback control has been implemented for NVC magnetometers, allowing the microwave frequency to be continuously adjusted to remain on resonance \cite{clevenson2018robust}. This allows for dramatic improvements in the dynamic range such that the magnetometers are principally limited by the microwave-source frequency-sweep range, the bias-field magnitude, and the NVC energy-level anticrossing point at approximately $\pm$ 100 mT \cite{doherty2013nitrogen}. 

\section*{Appendix L: Orientation Dependence}

\begin{figure}[h!]
\includegraphics[width=\columnwidth]{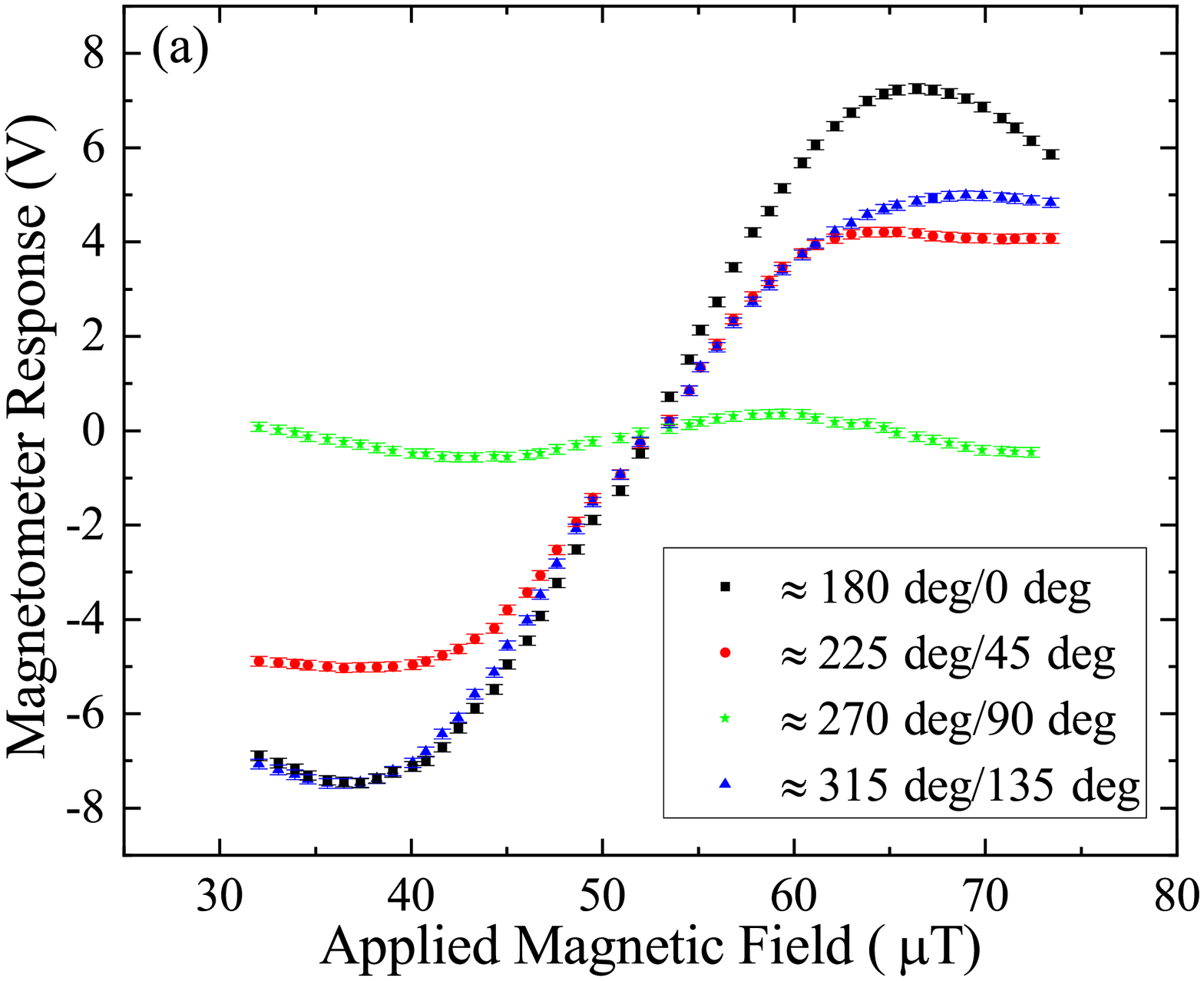}
\includegraphics[width=\columnwidth]{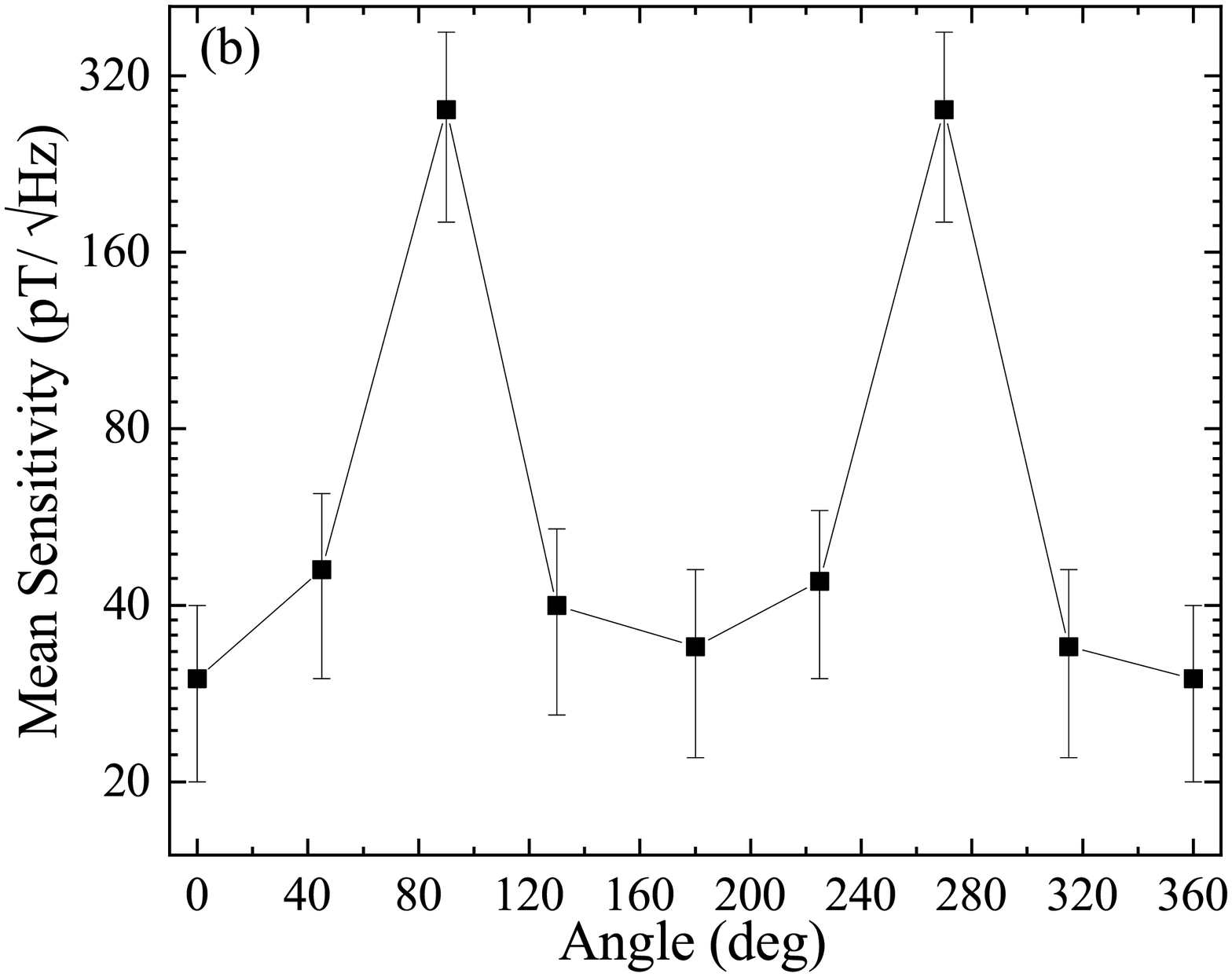} 
\caption{\small (a) The magnetometer response as a function of the applied test field, with the test field applied along different angles relative to the typical [100] orientation used in the main text.
(b) The mean sensitivity, taken for a (10 - 500)-Hz frequency range, using calibration constants determined from the magnetometer response to test fields applied along various directions in a 360 $^{\circ}$ circle about the sensor head. The discrepancy in mean sensitivity between 0 $^{\circ}$ and 180 $^{\circ}$ is likely due to a slight misalignment of the Helmholtz coil.} 
\label{fig: OrientationFFTs}
\end{figure}

NVCs only measure the projection of external magnetic fields along their symmetry axis, as opposed to the total magnetic field. As a result, the magnetometer responsitivity is weaker for magnetic fields applied orthogonally to a given NVC symmetry axis. This means that the ability of the magnetometer to detect magnetic signals depends not only on distance but also on the orientation of the sample relative to the diamond (and thus the NVC ensemble). In our case, the sensitive axis is not one of the $\langle$111$\rangle$ NVC symmetry axes but, instead, a vector along a [100] crystallographic orientation that projects equally onto each of the NVC symmetry axes. Accordingly, our NVC magnetometer is most sensitive to fields applied along this orientation and, as noted previously, the NVCs will not experience the total field but the projection of this total field along each of their symmetry axes and thus a field applied along [100] will always be reduced by the angle factor of 0.58 from the perspective of each NVC alignment. 
In contrast to the [111] bias-field-alignment case, however, there is no complete dead zone orthogonal to the sensitive axis, as the whole NVC population is being employed and there will always be some nonzero projection along one of the possible NVC symmetry axis alignments. However, the voltage response to the field is more complicated when the test field is not applied along [100], as the projection differs for each of the four NVC alignments. If the projection of an applied magnetic field can be measured for each possible NVC symmetry axis alignment it is then possible to obtain vector information \cite{barry2016optical}. Figure \ref{fig: OrientationFFTs}(a) shows the magnetometer response observed when a test field is applied along various directions. The stated angles are relative to the typical [100] orientation that we use, which is taken to be 0 $^{\circ}$ and 180 $^{\circ}$ for the two polarities of the Helmholtz coil. The exact orientation of the diamond lattice relative to the applied field is unknown. Figure \ref{fig: OrientationFFTs}(b) shows the mean sensitivity as a function of the angle, using calibration constants determined from the magnetometer-response plots of figure \ref{fig: OrientationFFTs}(a). This demonstrates the lack of dead zones in which the magnetometer is entirely insensitive. Even when the test field is applied orthogonal to the typical [100] direction a sensitivity of $\break$ (140 $\pm$ 50) pT/$\sqrt{\textrm{Hz}}$, in a (10 - 500)-Hz frequency range, is still observed.    

In practical terms, this orientation dependence means that, for example, with the turbo pump directly in front of the sensor head, it is possible to detect the 1-kHz signal at distances up to approximately 1.8 m away, but when the turbo pump is placed very roughly orthogonal to, and below, the sensor head, the 1-kHz signal can only be seen at up to $\break$ approximately 50 cm away. There could be some potential benefits to this orientational or directional dependence to magnetic signals, as the sensor head would be less sensitive to interference from certain noise sources, although we would have little control over the relative direction of magnetic fields within the environment and thus it would be more desirable to be able to measure the total magnetic field in general. For NVC ensemble magnetometers, it is possible to achieve this using a single diamond sample, if the bias field is aligned such that all eight possible ODMR resonances do not overlap. This would entail a reduction in contrast compared to a [100] orientation but would enable us to measure the projection of the sample magnetic field along the four different NVC alignments and from this, it would be possible to determine the $\textit{x}$, $\textit{y}$, and $\textit{z}$ magnetic field components and from these the total magnetic field. Vector magnetometry of this kind has been demonstrated successfully \cite{clevenson2018robust, schloss2018simultaneous, chatzidrosos2021fiberized}. It would be possible to measure the projection of the magnetic field along each NVC symmetry axis alignment with a [100] bias field; however, the applied field would need to be sufficiently large to separate out the resonances of each NVC alignment so that they are no longer completely overlapped.

\FloatBarrier

\section*{Appendix M: Electric Scooter Measurements}

\begin{figure}[h!]
\includegraphics[width=\columnwidth]{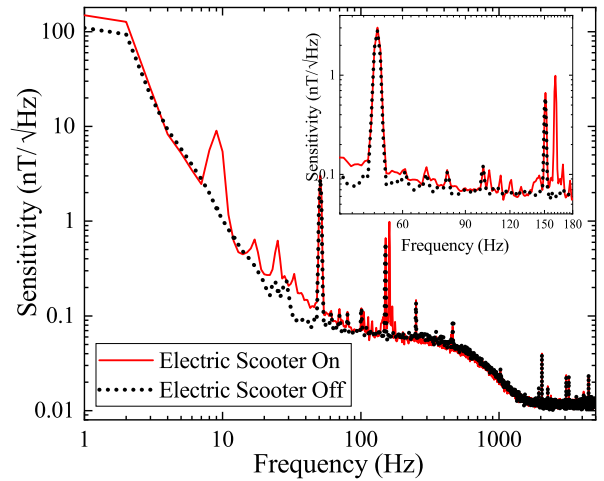} 
\caption{\small Sensitivity measurements taken with and without the electric scooter being turned on. Note that ac coupling is not used for these measurements. The sampling rate is set to 10 kHz. 
} 
\label{fig: Scooter1}
\end{figure}

Electric scooters have grown in popularity over recent years, being relatively environmentally friendly, low in cost, and increasingly accessible \cite{trivedi2019injuries}. Using our NVC magnetometer it is possible to detect various signals from the spinning back wheel, which contains the motor, 
of a Voi electric scooter, as can seen in Fig. \ref{fig: Scooter1}. In this case, the back wheel of the scooter (and the motor) is $\break$ approximately 92 cm from the diamond located within the sensor head. Unlike the vacuum-pump measurements, the scooter is not level with the sensor head and is instead placed on the ground.

These results, alongside the vacuum-pump measurements, suggest the potential use of NVC magnetometers for the condition monitoring of vehicles and more generally for remotely monitoring the usage and condition of machinery, even through thick walls. Failure modes could be identified from changes to the magnetic signals that appear in the FFT when compared to a working-condition reference signal set. This form of fault detection can be automated and does not require a detailed understanding of the device being investigated \cite{wang2006condition, verma2020automatic}.

\section*{Appendix N: Laser Noise}

\begin{figure}[h!]
\includegraphics[width=\columnwidth]{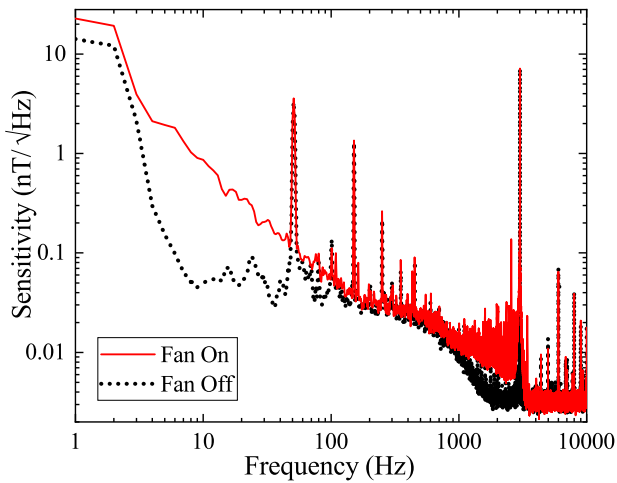} 
\caption{\small Sensitivity spectra taken with the laser-head fan turned on and off. No ac coupling is used for these measurements.} 
\label{fig: FanOnvsOff}
\end{figure}

Due to limited space and the consequently small size of the laser-head heat sink, it is necessary to cool the laser using active measures such as a fan. 
The fan is turned off when taking measurements and with the current laser power, the laser is not in a thermally stable regime. This may be responsible for at least some of the noise at low, sub-10-Hz, frequencies, the elevated and changing temperature of the laser head causing the laser to oscillate between different modes, producing large changes in laser output power. The use of a larger heat sink, or a lower-noise active solution, would enable the laser to operate at higher laser powers (important for pulsed magnetometry) and in a more stable thermal regime. The fan, as can be seen in Fig. \ref{fig: FanOnvsOff},
causes significant noise especially at lower frequencies due to the vibration of the optics, the kicking up of dust that passes through the laser beam, and the effect of temperature on components such as the PBS, as well as the magnetic signal from the fan.

Another factor to consider is the use of optics such as the PBS and optical isolator. It appears that the polarization of the laser output is unstable and the use of such polarization-maintaining optics may be causing large changes in the total amount of laser power being delivered to the diamonds and, in the case of the PBS, changing the ratio of the power delivered between the signal and reference sensor heads respectively. The noise caused by the latter effect cannot be cancelled using the balanced detector. Laser-performance issues in non-temperature-controlled environments are likely to be problematic for the practical implementation of high-sensitivity NVC magnetometers.

\FloatBarrier



%

\end{document}